\documentclass[fleqn,usenatbib]{mnras}

\usepackage{newtxtext,newtxmath}

\usepackage[T1]{fontenc}

\usepackage[normalem]{ulem}

\usepackage{verbatim}

\DeclareRobustCommand{\VAN}[3]{#2}
\let\VANthebibliography\thebibliography
\def\thebibliography{\DeclareRobustCommand{\VAN}[3]{##3}\VANthebibliography}

\usepackage{graphicx}	
\usepackage{amsmath}	
\usepackage{hyperref}
\usepackage{lipsum}
\usepackage{csquotes}
\usepackage{multicol}
\usepackage{multirow}
\usepackage{booktabs}

\title[\text{Inhomogeneous GCE Models for SNe Ia Contributions}]{Constraining SN Ia Progenitors from the Observed Fe-peak Elemental Abundances in the Milky Way Dwarf Galaxy Satellites}

\author[Alexander and Vincenzo]{
Ryan K. Alexander$^{1,3,4}$\thanks{E-mail: r.alexander-2021@hull.ac.uk}
and Fiorenzo Vincenzo$^{2,4}$
\\ ~ \\
$^{1}$E.A. Milne Centre for Astrophysics, University of Hull, Cottingham Road, Kingston-upon-Hull, HU6 7RX, United Kingdom\\
$^{2}$Dipartimento di Fisica e Astronomia “Ettore Majorana”, Università degli Studi di Catania, Via S. Sofia 64, 95123 Catania, Italy\\
$^{3}$Centre of Excellence for Data Science, AI, and Modelling (DAIM), University of Hull, Cottingham Road, Kingston-upon-Hull, HU6 7RX, United Kingdom\\
$^{4}$Joint Institute for Nuclear Astrophysics, Center for the Evolution of the Elements (JINA-CEE)
}

\date{Accepted XXX. Received YYY; in original form ZZZ}

\pubyear{2024}

\begin{document}
\label{firstpage}
\pagerange{\pageref{firstpage}--\pageref{lastpage}}
\maketitle

\begin{abstract}

\noindent Chemical abundances of iron-peak elements in the red giants of ultra-faint dwarf galaxies (UFD) and dwarf spheroidal galaxies (dSph) are among the best diagnostics in the cosmos to probe the origin of Type Ia Supernovae (SNe Ia). We incorporate metallicity-dependent SN Ia nucleosynthesis models for different progenitor masses in our inhomogeneous galactic chemical evolution model, {\tt{i-GEtool}}, to recreate the observed elemental abundance patterns and their spread in a sample of UFD and dSph galaxies with different average metallicities and star formation histories. Observations across different environments indicate that both [Ni/Mg] and [Mn/Mg] increase on average with metallicity, with the latter having a higher slope. The average dispersion of [X/Mg] from our UFD model ranges between $0.20$ and $0.25$ for iron-peak elements, with the exception of [Sc/Mg] that has $\sigma \approx 0.39$. Chemical evolution models assuming Chandrasekhar mass (M$_{\text{ch}}$) SN Ia progenitors produce similar [Ni/Mg]-[Fe/H] and [Mn/Mg]-[Fe/H] abundance patterns to those observed in the examined UFD and dSph galaxies, without the need to invoke a substantial fraction of sub-M$_{\text{ch}}$ progenitors that change across different environments, as claimed by some previous chemical evolution studies. Sub-M$_{\text{ch}}$ progenitors in our dSph models under produce both [Ni/Mg]-[Fe/H] and [Mn/Mg]-[Fe/H] abundance patterns. We stress on the importance of accounting for inhomogeneous chemical enrichment and metallicity-dependent SN Ia yields, which are the main aspects that distinguish our work from the previous chemical evolution studies of iron-peak elements.


\end{abstract}





\section{Introduction}

Type Ia Supernovae (SNe Ia) are among the most dominant events contributing to galaxies' chemical evolution. Such events are the primary contributors to the origin of iron-peak elements in the Universe \citep{tinsley1979,matteucci1985,matteucci1986}. According to the main proposed scenarios, they occur through the thermonuclear explosion of an electron-degenerate carbon-oxygen (CO) white dwarf (WD), which interacts with a companion star in a binary system until the WD attains central densities $\rho_{c}\approx 2 \times 10^{9}\,\text{g}\,\text{cm}^{-3}$ and temperatures $T_{c}\approx 2 \times 10^{8}\,\text{K}$, corresponding to masses close to the Chandrasekhar mass (M$_{\text{ch}}$). At those densities and temperatures, carbon can be ignited in the electron-degenerate core from which nuclear runaway reactions occur and the WD eventually explodes 
(e.g., see \citealt{arnett1996,rauscher2020}). 

Two main scenarios have been postulated to achieve ignition condition, either via the accretion from a binary star companion onto the WD by Roche lobe overflow (single degenerate scenario; see \citealt{whelan1973,nomoto1984,thielemann1986,hachisu2012}) or via the merger between two CO WDs (double degenerate scenario; see \citealt{iben1984,webbink1984}). 
It has been proposed that either channel can give rise to a double detonation of the WD at sub-Chandrasekhar masses (sub-M$_{\text{ch}}$) \citep{woosley1994}; in particular, the WD can accrete He-rich material from the companion until a detonation takes place in the He shell, which causes the development of shock waves that propagate through the WD and reach the core, where a secondary detonation occurs as carbon burning ignites \citep{fink2007,fink2010,woosley2011,floers2020,gronow2021}.

The elemental abundance templates that result from the thermonuclear explosion of a SN Ia are expected to depend on the birth metallicity of the WD progenitor in the binary system (e.g., see \citealt{hoflich1998,timmes2003}). In particular, the amounts of C, N and O at birth in the WD progenitor regulate how much {$^{14}$N} is synthesized in the CNO cycle. {$^{14}$N} is then used during He-burning to produce {$^{22}$Ne} from a series of $\alpha$-captures interlaced with a $\beta$-decay. The amount of {$^{22}$Ne} in the CO core of the WD progenitor is critical because it determines the neutron excess, which is among the key factors in regulating the abundances of the synthesized nuclides under nuclear statistical equilibrium (e.g., see \citealt{hartmann1985}). Since the amount of {$^{22}$Ne} in the WD depends on the C, N, and O abundances at birth in the star, different birth metallicities give rise to different SN Ia nucleosynthetic yields (e.g., see \citealt{howell2009,townsley2009}). The elemental abundance template from a SN Ia explosion also depends on the mass and density of the WD, with Mn and Ni being among the chemical elements with the largest changes (e.g., see \citealt{seitenzahl2013,lach2020}). 

In this work, we aim to investigate the effects of metallicity-dependent M$_{\text{ch}}$ and sub-M$_{\text{ch}}$ SN Ia models on the chemical evolution of a sample of ultra-faint dwarf galaxies (UFD) and dwarf spheroidal galaxies (dSph) that are satellites of the Milky Way (MW), by making use of {\tt{i-GEtool}}, an inhomogeneous chemical evolution model \citep{alexander2023}.  Various works in the literature aimed at constraining the SN Ia progenitors from the observed iron-peak elemental abundances in stars. An interesting study is that of \citet{seitenzahl2013}, who tested various SN Ia models to constrain the fraction of near-M$_{\text{ch}}$ to sub-M$_{\text{ch}}$ SNe Ia in the Galaxy. They found near-M$_{\text{ch}}$ SNe Ia are needed to reproduce the observed [Mn/Fe] in the Solar neighbourhood at [Fe/H] $\geq$ 0, with 50\% of SN Ia events stemming from near-M$_{\text{ch}}$ progenitors. \citet{seitenzahl2013} also proposed that a different combination of sub-M$_{\text{ch}}$ and M$_{\text{ch}}$ SNe Ia could be a possible solution to explain [Mn/Fe] in dSphs. The possibility of different SN Ia progenitors in dSphs was explored by \citet{mcwilliam2018}, who aimed to explain the observed elemental abundances of a star in the ancient dSph galaxy Ursa Minor. \citet{mcwilliam2018} determined that the observed abundances in the star result from a diluted chemical enrichment from a single sub-M$_{\text{ch}}$ SN Ia progenitor. The work of \citet{mcwilliam2018} was further expanded by \citet{kirby2019}, who investigated the chemical evolution of dSph galaxies with different star formation histories (SFHs). \citet{kirby2019} found that the observed [Ni/Fe] in ancient and short-lived dSphs (like Ursa Minor dSph in  \citealt{mcwilliam2018}) are compatible with a pure chemical enrichment of only sub-M$_{\text{ch}}$ SN Ia progenitors, while the observed [Ni/Fe] in the MW and dSphs with more extended SFHs indicate that the chemical enrichment was contributed by different classes of SNe Ia.

While \citet{kirby2019} focused on constraining the SN Ia progenitors in dSphs from the observed Ni abundances, \citet{delosreyes2020} addressed a similar question by using a different iron-peak element, Mn. In particular, \citet{delosreyes2020} measured Mn abundances in a sample of dSph galaxies, increasing the available sample size towards lower metallicities in the literature. The results of \citet{delosreyes2020} based on the [Mn/Fe] distribution in dSphs agree with those of \citet{kirby2019} based on [Ni/Fe]. Interestingly, \citet{delosreyes2020} suggested that the different [Mn/Fe] in dSphs with different SFHs could be explained if M$_{\text{ch}}$ SN Ia progenitors occurred on longer average delay times than sub-M$_{\text{ch}}$ from the formation of the binary system. Another interesting work that also determined a larger contribution of M$_{\text{ch}}$ SN Ia progenitors in galaxies with more extended SFH is that of \citet{kobayashi2020}, who incorporated several metallicity-dependent SN Ia models into a detailed chemical evolution model, finding that the observed [Ni/Fe] and [Mn/Fe] in the Solar neighbourhood require up to 25 per cent of sub-M$_{\text{ch}}$ SNe Ia progenitors, with a higher percentage of sub-M$_{\text{ch}}$ being present for dSph galaxies. More recently, \citet{delosreyes2022} developed detailed chemical evolution calculations to analyse the star formation and chemical enrichment histories of the ancient dSph galaxy, Sculptor, fitting the observed abundance patterns of several chemical elements with their detailed model. In agreement with their previous works on the subject, \citet{delosreyes2022} determined that the observed [Ni/Fe] and [Mn/Fe] in Sculptor point towards a scenario in which sub-M$_{\text{ch}}$ SN Ia progenitors likely dominated in the chemical evolution of iron-peak elements within this galaxy. Other works in the literature who drew similar conclusions are \citet{sanders2021} and, more recently, \citet{nissen2024}. 



A work in the literature that points towards a different scenario with respect to \citet{kobayashi2020} is that of \citet{eitner2020}, who determined that the observed [Mn/Fe] in the MW are reproduced by assuming $\approx 75$\% of sub-M$_{\text{ch}}$ SN Ia progenitors, with the remaining fraction stemming from M$_{\text{ch}}$ SNe Ia. \citet{gronow2021} also found that sub-M$_{\text{ch}}$ SNe Ia along with core-collapse Supernovae (CCSNe) account for more than 80 per cent of solar Mn when considering \citet{limongi2018} massive star yields, in agreement with the findings of \citet{palla2021}, who performed one of the most systematic investigations on the effect of different SN Ia nucleosynthesis models to explain the observed chemical abundance patterns of MW stars. 

Our work is unique among previous literature in this field as our methodology is based on an inhomogeneous chemical evolution model that incorporates metallicity-dependent SN Ia yields for different WD progenitor masses. The purpose of our work is to investigate the chemical abundances of Fe-peak elements for varying SN Ia yields and to constrain the elemental contributions from SN Ia progenitors, exploring the predictions of stellar
nucleosynthesis calculations of exploding white dwarfs with different masses and metallicities. Our work is organised as follows. Section \ref{section:type1a_yields} provides a detailed description of the metallicity-dependent SNe Ia yields we use in our work. The chemical evolution models are presented in Section \ref{section:models}, where we discuss some of the physical parameters of the study. Section \ref{section:observational} summarises our observational samples with a full recollection of the data. We present our results in Section \ref{section:results}, where we discuss the predicted chemical abundance ratios when varying metallicity-dependent yields. Lastly, we show the conclusions of our work in Section \ref{section:conclusions}. 

\begin{figure*}
    \centering
    \includegraphics[width = 1.0\textwidth]{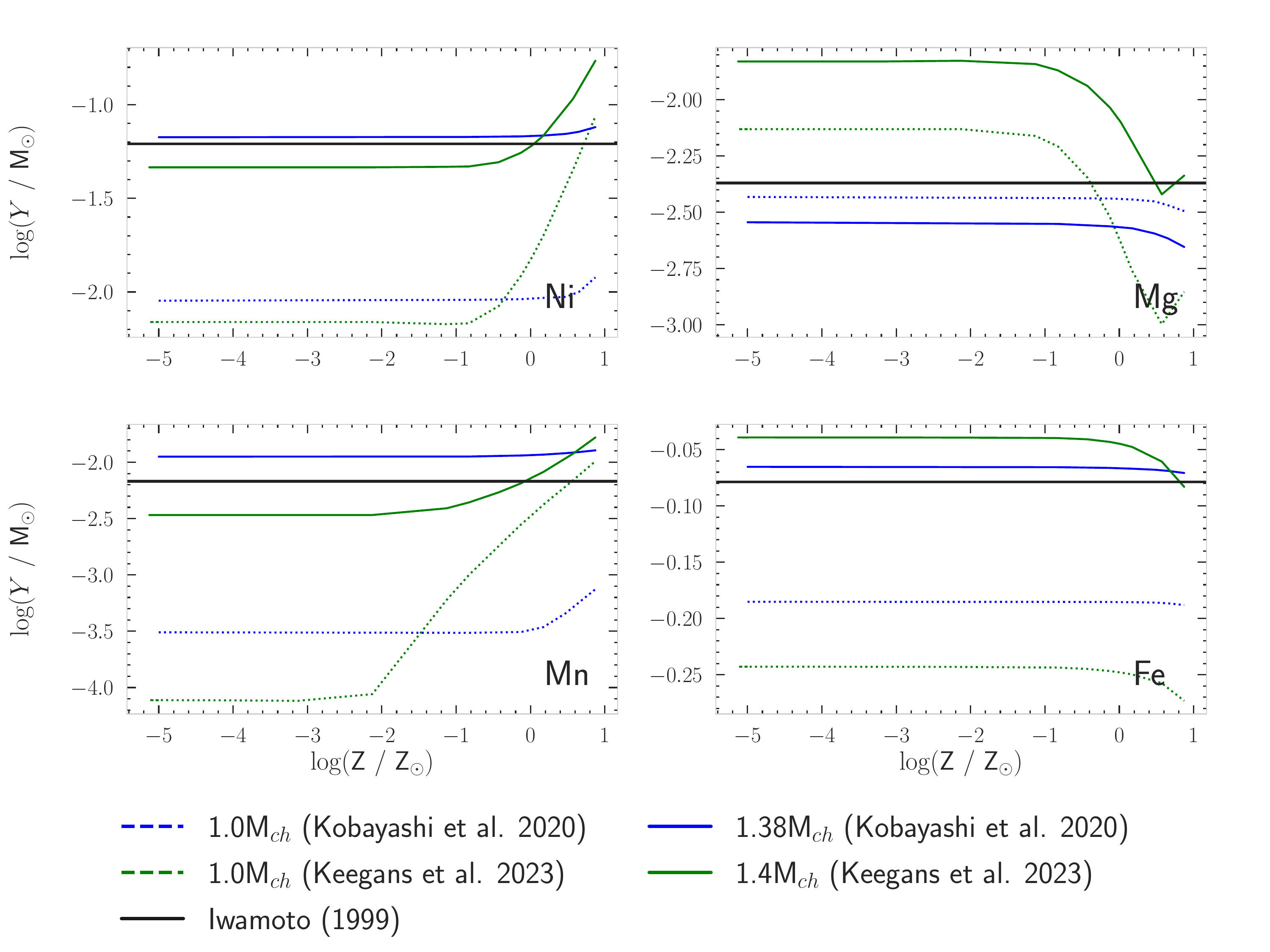}
    \caption{The SN Ia nucleosynthesis yields of \citet{kobayashi2020} and \citet{keegans2023} for Ni (upper left), Mg (upper right), Mn (bottom left) and Fe (bottom right) as a function of metallicity. Various progenitors are shown through colours where the blue lines represent \citet{kobayashi2020} WD progenitors with masses $1.0 \,\text{M}_{\sun}$ (dotted) and $1.38 \,\text{M}_{\sun}$ (solid), respectively, and the green lines represent \citet{keegans2023} WD progenitors with masses $1.0 \,\text{M}_{\sun}$ (dotted) and $1.4 \,\text{M}_{\sun}$ (solid), respectively. The black solid line corresponds to the W7 non-metallicity dependent SN Ia yields of \citet{iwamoto1999}.}
    \label{fig:yields_comp}
\end{figure*}



\begin{figure*}
    \centering
    \includegraphics[width = 1.0\textwidth]{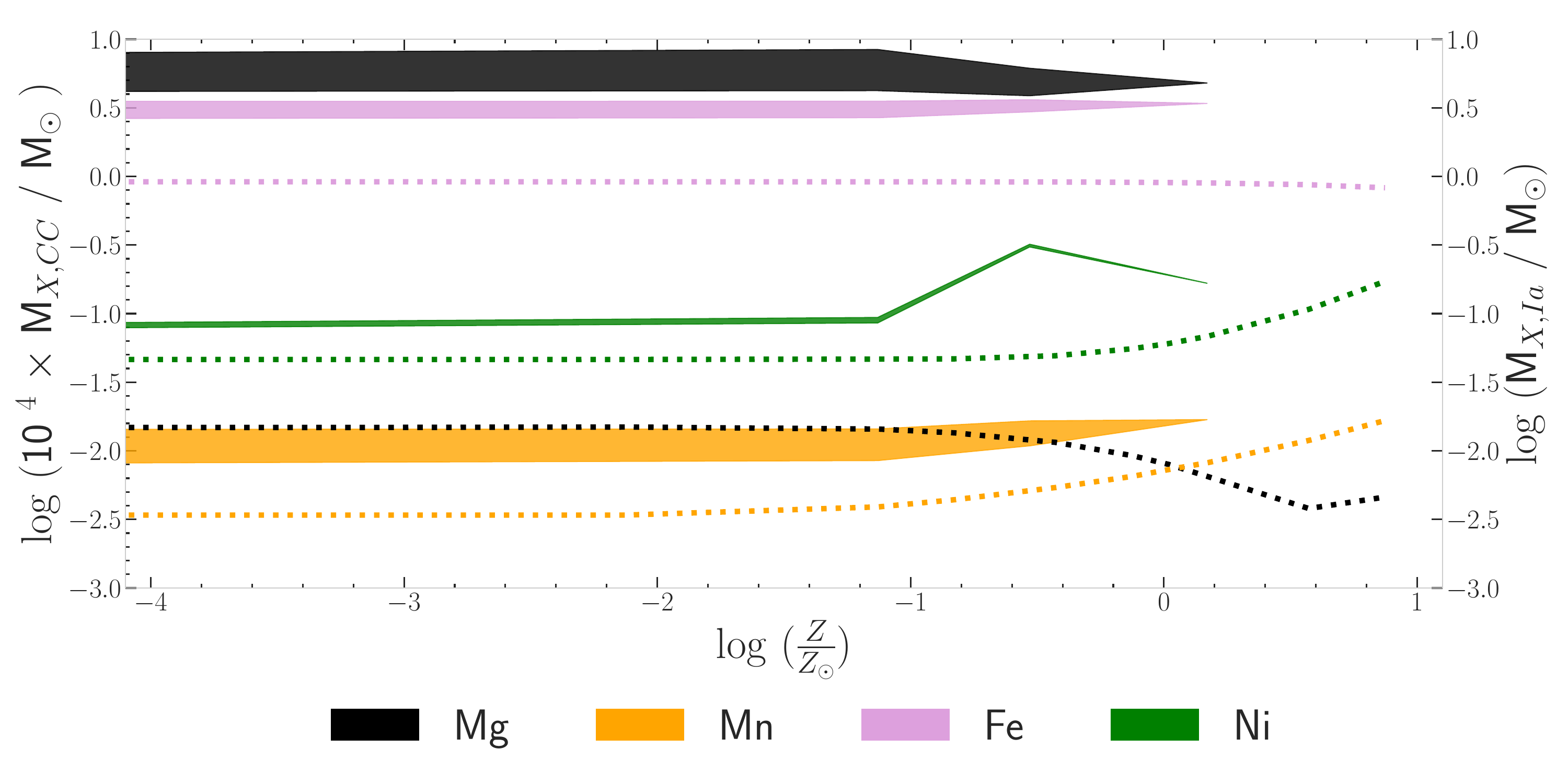}
    \caption{A comparison between the total massive star yields of \citealt{kobayashi2006} for a $10^{4}\,\text{M}_{\sun}$ stellar population (left $y$-axis, filled bars) and the \citet{keegans2023} yields from a single M$_{\text{ch}}$ SN Ia event with progenitor mass $M_{\text{WD}}=1.4\, \text{M}_{\sun}$ (right $y$-axis, dotted lines). The nucleosynthesis yields of Mg are shown in black, Mn in orange, Fe in pink and Ni in green. For each element, the width of the bar corresponds to the total massive-star contribution when changing the massive star cutoff for CCSNe from $25\,\text{M}_{\sun}$ (lowest edge of the bar) to $40\,\text{M}_{\sun}$ (highest edge).}
    \label{fig:snia_ccsn_comp}
\end{figure*}

\section{Type Ia Supernova Nucleosynthetic Yields}
\label{section:type1a_yields}

In this section, we present the SN Ia models that are adopted in our chemical evolution calculations with {\tt{i-GEtool}}. The section is split into two parts. The first part describes the SN Ia models of \citet{kobayashi2020}, while the second part reviews those of \citet{keegans2023}; both SN Ia yields are compared with the W7 SN Ia yields of \citet{iwamoto1999}, widely assumed by chemical evolution calculations in the literature \citep{kobayashi2006,romano2010,matteucci2012,prantzos2018,womack2023}. 


\subsection{Type Ia Supernova models of Kobayashi et al. (2020)}

\cite{kobayashi2020} developed SN Ia models for a variety of WD progenitor masses and metallicities by using the 2D hydrodynamical code of \citet{leung2015}. To have a larger number of isotopes with associated yields for their chemical evolution calculations, \cite{kobayashi2020} post-processed the nucleosynthesis with a tracer particle scheme \citep{travaglio2004}, using the so-called \textit{Torch} nuclear reaction network \citep{timmes1999}. SN Ia models and yields were calculated for WD progenitor masses $M_{\text{WD}}=0.9$, $1.0$, $1.1$, $1.2$, $1.3$, $1.33$, $1.37$, and $1.38$ M$_{\sun}$ at the following metallicities: $Z = 0$, $0.002$, $0.01$, $0.02$, $0.04$, $0.06$ and $0.1$. 
Here, we adopt the two benchmark models of \cite{kobayashi2020}, that have WD progenitor masses $M_{\text{WD}}=1.0$ M$_{\sun}$ (sub-M$_{\text{ch}}$ double detonation model) and $M_{\text{WD}}=1.38$ M$_{\sun}$ (near-M$_{\text{ch}}$ deflagration detonation model), used also in their chemical evolution calculations for comparison against observational data in the MW and dSph galaxies. 

In Fig. \ref{fig:yields_comp}, we show how the total yields of Mn, Fe, Ni, and Mg change as a function of metallicity according to the benchmark models of \citet{kobayashi2020}. We also include the W7 SN Ia yields of \citet{iwamoto1999} for each of the elements as a baseline (horizontal black lines), which are constant as a function of metallicity. As our current work focuses on Fe-peak elements, we choose to analyse Ni, Mn and Fe as they are predominantly created in SNe Ia events. We also include Mg as it is an $\alpha$-element mainly produced by CCSNe, typically used as a \enquote*{chemical clock} when its abundance is analysed relative to an iron-peak element in chemical evolution studies \citep{matteucci1986}. From the figure, higher SN Ia progenitor masses contribute more iron-peak elements to the ISM of a galaxy. Interestingly enough, the production of Ni is higher than Mn for all considered progenitor masses, including the W7 \citet{iwamoto1999} yields. Finally, for sub-solar metallicities, the SN Ia yields of \citet{kobayashi2020} exhibit a weak dependence on metallicity. Some change with metallicity starts appearing at super-solar metallicities, especially for Mn, which increases as $Z$ also increases.  

\subsection{Type Ia Supernova models of Keegans et al. (2023)}

\citet{keegans2023} post-processed the $0.8$ M$_{\sun}$ double detonation model of \citet{miles2019}, the $1.0$ M$_{\sun}$ double detonation model of \citet{shen2018} and the $1.4$ M$_{\sun}$ mass deflagration detonation model of \citet{townsley2016}. 
For each progenitor mass, \citet{keegans2023} considered the following metallicities: $Z = 0$, $10^{-7}$, $10^{-6}$, $10^{-5}$, $0.0001$, $0.001$, $0.002$, $0.005$, $0.01$, $0.014$, $0.02$, $0.05$ and $0.1$, developing a total number of 39 unique SN Ia models. The initial abundance of metals of the WD for each of these models is based on a uniform distribution of the mass fraction of $^{22}$Ne. 

Nucleosynthesis is post-processed by using the so-called \textit{Tracer Particle Post-Processing Network-Parallel} (`tppnp' for short) of the NuGrid collaboration \citep{pignatari2016,ritter2018,jones2019}. The adopted nuclear reaction network allowed \citet{keegans2023} to cover over 5,000 isotopes and over 70,000 reactions whereas the works of \citet{townsley2016}, \citet{shen2018} and \citet{miles2019} are more restricted in their reaction networks. 
The CO WD is initialised to be as close to the chemical make-up as seen in the original models, varying the $^{22}$Ne mass fraction. One key difference between the \citet{keegans2023} yields and the original is the under production of some of the heavy-iron group elements such as Cu, Ga and Ge. The majority of the other elements however are very similar to each other, except for N and Mg that have a higher production in \citet{keegans2023}. 

In addition to \citet{kobayashi2020} yields, Fig. \ref{fig:yields_comp} also shows the total yields of Mn, Fe, Ni, and Mg as predicted by the $1.0$ M$_{\sun}$ (dotted green line) and $1.4$ M$_{\sun}$ (solid green line) SNe Ia models of \citet{keegans2023} as a function of metallicity. We also show the W7 yields of \citet{iwamoto1999} as a baseline. The behaviour of the \citet{keegans2023} yields as a function of the SN Ia progenitor mass is similar to that of \citet{kobayashi2020}, as higher progenitor masses are predicted to produce a larger amount of iron-peak elements. At very low metallicities ($\text{[Fe/H]} \lesssim -2$), the SN Ia models of \citet{keegans2023} predict nearly constant yields as a function of $Z$, and the 1.4 M$_{\sun}$ model predicts similar iron-peak elemental yields as the W7 \citet{iwamoto1999} model. When the metallicity passes a certain threshold ($\text{[Fe/H]} \gtrsim -2$), all progenitor masses experience an increase in the production of iron-peak elements, with the 1.0 M$_{\sun}$ model exhibiting a stronger metallicity-dependence than the 1.4 M$_{\sun}$ model.

Massive stars dying as CCSNe also provide some contribution to iron-peak elements, even though they mainly contribute to $\alpha$-elements such as Mg. Fig. \ref{fig:snia_ccsn_comp} details the total contribution of Mg (red), Mn (green), Fe (blue) and Ni (black) from a population of massive stars formed out of a stellar population with total stellar mass $M_{\star}=10^{4}\,\text{M}_{\sun}$ and varying metallicity, by assuming the \citet{kobayashi2006} massive star yields and the IMF of \citet{kroupa2001} (left y-axis, filled bars), the same assumed in our calculations with {\tt{i-GEtool}}. The filled bars in the figure are computed by varying the upper mass-limit for the chemical enrichment of massive stars from $25\,\text{M}_{\sun}$ (lower edge of the bars) to $40\,\text{M}_{\sun}$ (upper edge of the bars). We choose a total stellar mass of $M_{\star}=10^{4} \, \text{M}_{\sun}$ for the stellar population as it is similar to the predicted mass of a UFD galaxy. In the same figure, we also show the total yields from the 1.4 M$_{\sun}$ model of \citet{keegans2023}, that correspond to the predicted chemical enrichment from a single SN Ia event (right y-axis, dotted lines). 

It is clear from Fig. \ref{fig:snia_ccsn_comp} that a single SN Ia event provides comparable amounts of iron-peak elements to those from massive stars in a $10^{4}$ M$_{\sun}$ stellar population. This highlights the significance of SNe Ia in the chemical enrichment and contamination of iron-peak elements in the ISM. We cannot ignore the differences between the nucleosynthesis yields of Mg from a single SN Ia and the CCSNe from a stellar population with stellar mass $M_{\star}=10^{4}\,\text{M}_{\sun}$  as the latter outweigh the former by $\approx 2.5$ orders of magnitude.
When increasing the cut-off mass for CCSNe from 25 M$_{\odot}$ to 40 M$_{\odot}$, the differences in the massive star contribution are larger for Mg, Mn and Fe while Ni does not change significantly. We also note that, as the metallicity approaches the solar value, the differences overall diminish.


\section{Inhomogeneous Chemical Evolution Models}
\label{section:models}

\begin{figure}
    \centering
    \includegraphics[width = 0.4\textwidth]{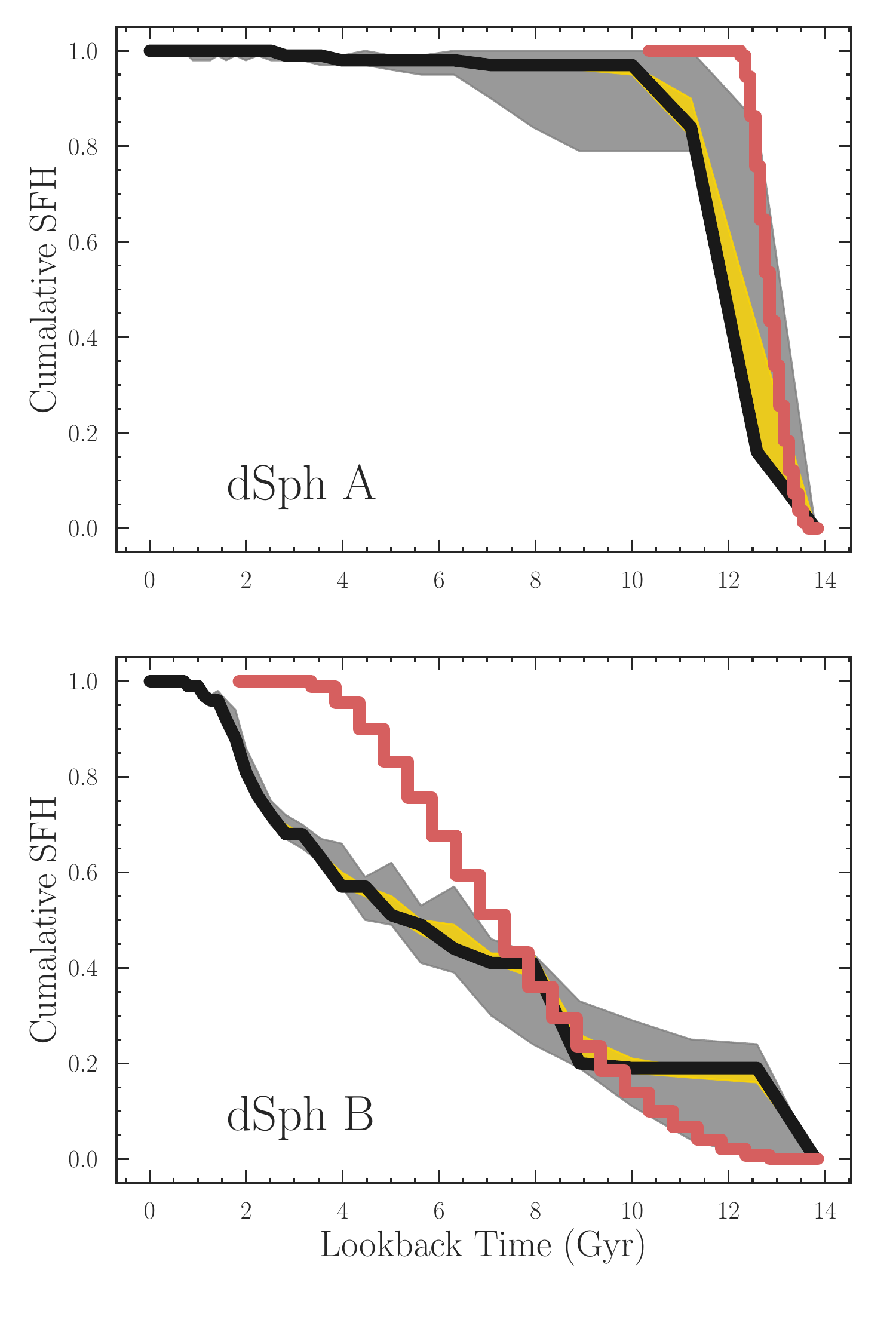}
    \caption{A comparison between the model cumulative star formation history (red solid lines) and the predicted best fit star formation histories (black solid lines) from \citet{weisz2014},  who fitted the observed colour-magnitude diagrams from deep HST photometry, for Sculptor (top panel) and Leo I (bottom panel), as a function of lookback time. In addition to these cumulative distributions from \citet{weisz2014}, we also show their 68\% interval from total uncertainty (grey) and random uncertainty (yellow).}
    \label{fig:sfh}
\end{figure}

$\tt{i-GEtool}$ is an inhomogeneous galactic chemical evolution tool which tracks the elemental abundances of stars with different masses, metallicities and ages along with the chemical makeup of the local ISM of galaxies as a function of time (see \citealt{alexander2023} for more details). We use $\tt{i-GEtool}$ to investigate how the assumption of metallicity-dependent SN Ia yields affects the chemical evolution of UFDs and dSphs. We create one UFD and two dSph toy models to cover different SFHs and metallicity ranges, from -$4 \lesssim$ [Fe/H] $\lesssim$ -2 of UFDs to [Fe/H] $\gtrsim$ -2 of dSphs. One dSph model aims to reproduce the chemical abundances of Sculptor-like dwarf galaxies with a short but intense SFH (henceforth, \textit{dSph A}) whereas the other is set up to have a long but shallow SFH (henceforth, \textit{dSph B}), similar to the inferred SFH of Leo I (e.g., see \citealt{weisz2014}).


\begin{figure*}
    \centering
    \includegraphics[width = 1.0\textwidth]{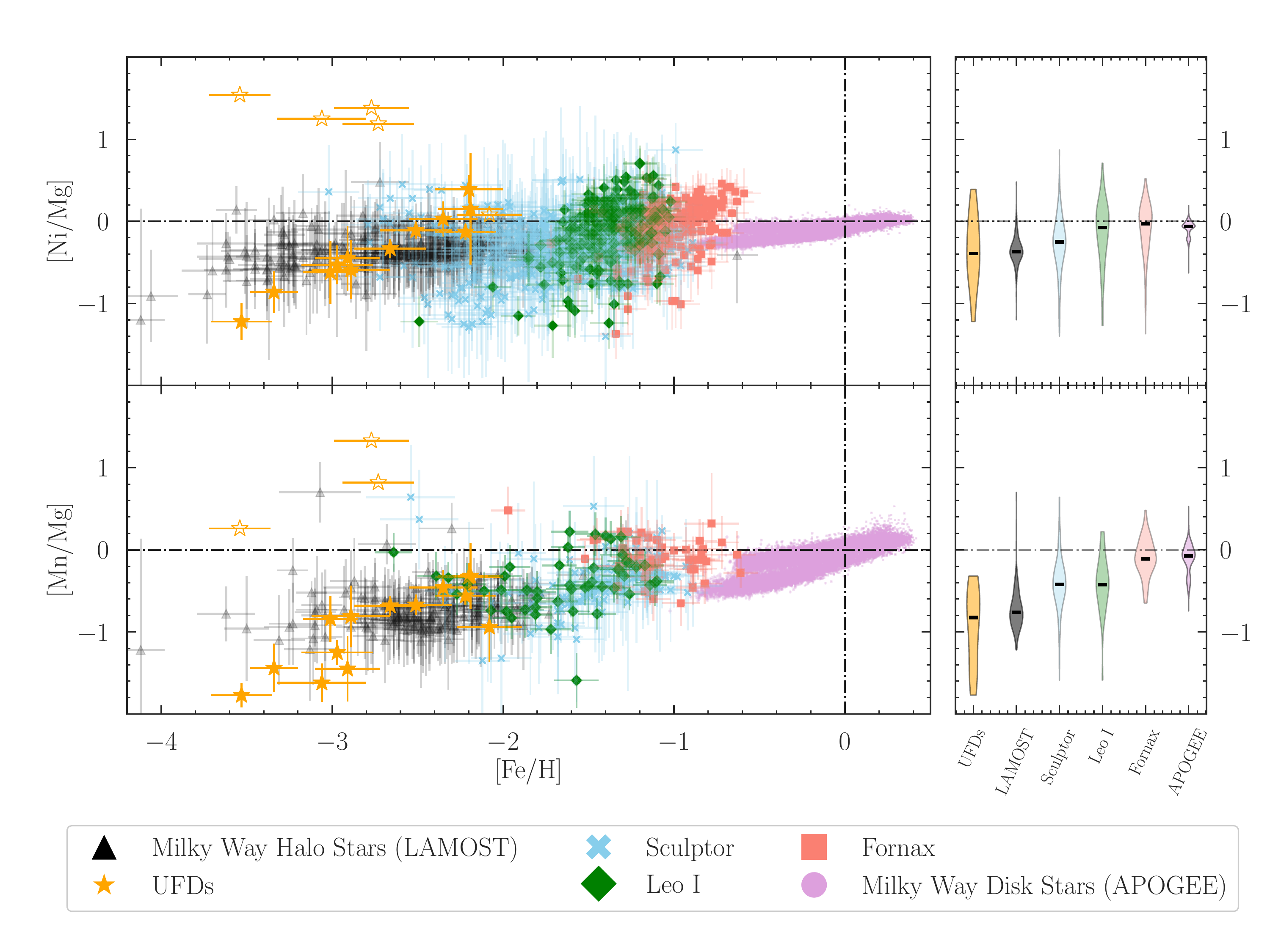}
    \caption{The observed [Ni/Mg]-[Fe/H] (top left panel) and [Mn/Mg]-[Fe/H] (bottom left panel) for MW halo stars (LAMOST - black), UFD stars (orange), Sculptor stars (blue), Leo I stars (green), Fornax stars (pink) and MW field stars (APOGEE - purple).The empty symbols correspond to upper limits. On the right are violin plots describing the distribution function of [Ni/Mg] (top right panel) and [Mn/Mg] (bottom right panel) in the different galaxies ordered from left to right according to their average stellar metallicity, with the black lines showing the mean of [X/Mg] within each sample.}
    \label{fig:snia_obs}
\end{figure*}

\textit{Initial setup ---} The simulation takes place within the constraint of a three-dimensional box where we place primordial gas of H and He with abundances $X = 0.74$ and $Y = 0.26$ in mass, respectively. These ratios are chosen to reproduce the conditions of the early Universe from the epoch of reionization. The box is divided into 40$^{3}$ cells, where the size of the cell is 20 pc for the UFD model and 40 pc for \textit{dSph A} and \textit{dSph B}. 
The models assume the following initial gas mass density, $\rho_{\text{gas}}(r,t=0)$, as a function of radius, $r$: 
\begin{equation}
    \rho_{\text{gas}}( r, t=0 ) = \rho_{0} \, b^{ - \frac{ r }{ 200\,\text{pc} } },
\end{equation}
\noindent where $\rho_{0}=55\,\text{M}_{\sun}\,\text{pc}^{-3}$ and $b=85$, giving rise to a total initial gas mass $M_{\text{gas,tot}}(t=0) = 3 \times 10^{6}\,\text{M}_{\sun}$ when the model is run.

\textit{Gas accretion ---} The SFH is driven by primordial gas accretion, where the inflow is uniformly distributed throughout the box and serves as a driving factor for star formation and subsequent chemical enrichment. The accretion rate has the same parametrization as in \citet{kobayashi2020} as follows:

\begin{equation}
    \Big( \frac{\text{d}\,\rho_{\text{gas}}(t)}{\text{d}t} \Big)_{\text{infall}} = \mu \times t \times \exp\Big(-\frac{t}{\lambda}\Big),
    \label{infall}
\end{equation}

\noindent where $t$ is the simulation time, $\mu$ measures the intensity of the gas accretion and $\lambda$ varies the duration of the gas accretion rate. The total gas mass accreted into the galaxy is $M_{\text{UFD, inf}} = 6\times10^{6} \, \text{M}_{\sun}$, $M_{\text{dSph A, inf}} = 4\times10^{8} \, \text{M}_{\sun}$, and $M_{\text{dSph B, inf}} = 9\times10^{8} \, \text{M}_{\sun}$ for the UFD model, \textit{dSph A} and \textit{dSph B}, respectively, where these values are a constraint on $\mu$. Table \ref{table:parameters} contains the values used for $\lambda$ and $\mu$ in our models.

\begin{table}

    \begin{tabular}{|p{2.4cm}|p{1.5cm}|p{1.5cm}|p{1.5cm}|}

        \hline

        \multicolumn{4}{|c|}{Free Parameters} \\

        \hline

        Parameters & UFD & dSph A & dSph B \\

        \hline

        Volume [kpc$^{3}$]: & 0.512 & 4.096 & 4.096 \\ 

        Infall Gas [M$_{\sun}$]: & 6$\times$10$^{6}$ & 4$\times$10$^{8}$ & 9$\times$10$^{8}$ \\

        $\lambda$ [Myr]: & 100 & 250 & 2000 \\

        $\epsilon$ [Gyr$^{-1}$]: & 0.004 & 0.20 & 0.04 \\

        \hline

        \multicolumn{4}{|c|}{Stellar Properties} \\

        \hline

        Properties & UFD & dSph A & dSph B \\

        \hline

        Number of Stars & 8.4$\times$10$^{4}$ & 2.4$\times$10$^{7}$ & 5.3$\times$10$^{7}$ \\

        Stellar Mass [M$_{\sun}$]: & 3.2$\times$10$^{4}$ & 9.2$\times$10$^{6}$ & 1.7$\times$10$^{7}$ \\

        SFH [Gyr]: & 1.9 & 2.0 & 11.0 \\

        \hline

    \end{tabular}

    \caption{Free Parameters of each model and stellar properties shown afterwards as we consider two different sets of metallicity-dependent yields for SNe Ia with different progenitor masses. The table highlights some of the differences between each model, with rows corresponding to UFD, dSph A and dSph B. \textit{Row 1}: Total volume of the box for each model. \textit{Row 2}: The total accreted gas in each model. \textit{Row 3}: The peak accretion of each model as a function of time. \textit{Row 4}: The star formation efficiency for each model. \textit{Row 5}: The total number of stars in each model. \textit{Row 6}: The stellar mass of each model at the point where star formation is truncated. \textit{Row 7:} The duration of the star formation activity.}

    \label{table:parameters}

\end{table}

\textit{Star formation ---} The star formation rate (SFR) is assumed to follow a linear Schmidt-Kennicutt relation \citep{schmidt1959, kennicutt1998}, $\text{SFR}(t) = \epsilon \times \big\langle \rho(t) \big\rangle \, V$, 
where $\epsilon$ is the star formation efficiency (SFE; see Table \ref{table:parameters}), $\big\langle \rho(t) \big\rangle$ is the average gas mass density and V is the total volume of the simulated galaxy. Our work assumed power-law index $n = 1$, similar to \citet{andrews2017} and \citet{cote2017}. \citet{vincenzo2017} compared the effect of linear and non-linear Schmidt-Kennicutt law in a chemical evolution model and found differences in the SFE and surface gas mass densities, which however lead to almost identical chemical abundance patterns.
At any given time step $\Delta$t, stars are formed from a random sampling of the \citet{kroupa2001} initial mass function (IMF) from $0.1\,\text{M}_{\sun}$ to $100\,\text{M}_{\sun}$. When star formation occurs within a cell, the star adopts the chemical composition of the adjacent eight cells (see \citealt{alexander2023} for more details). 


\textit{Stellar yields ---} In our model, we adopt the same stellar nucleosynthesis yields as in \citet{alexander2023}. In particular, we assume the yields of \citet{kobayashi2006} for high-mass stars along with the yields \citet{karakas2010} for asymptotic giant-branch (AGB) stars. At the end of the stars' lifetimes, they dump enriched material into the surrounding ISM for future stellar populations to inherit (see \citealt{alexander2023} for details). 


\textit{Delay-time distribution for Type Ia Supernovae ---} To model the delay-time between the formation of the SN Ia progenitor system and the SN Ia explosion, we assume 
a power law of the form, $\text{DTD}_{\text{Ia}}(\tau) = N_{\text{Ia}} \tau^{-1.1}$ which is motivated by a number of observational studies (e.g., see \citealt{maoz2010,maoz2014,maoz2017,castrillo2021}) and previous chemical evolution calculations (e.g., see \citealt{weinberg2024}). From several tests, Our toy models assume $N_{\text{Ia}} = 10^{-3} \, \text{M}^{-1}_{\sun}$, and a minimum delay-time $\tau_{\text{min}} = 150 \, \text{Myr}$. Our chosen normalization is in good agreement with previous works (e.g. see \citealt{maoz2010, maoz2014, graur2011, rybizki2018, weinberg2024} for more details), along with the minimum delay time (see \citealt{andrews2017} for more details).

\textit{Outflows ---} All supernovae events are assumed to have an identical explosion energy, E$_{\text{SN}}$ = 10$^{51}$ ergs, sweeping out M$_{\text{swept}}$ = 5 $\times$ 10$^{4}$ M$_{\sun}$ of the ISM mass. Galactic outflows are driven by SN explosions in the form of a SN bubble, where the swept-up mass M$_{\text{swept}}$ is lost from the galaxy if any part of the bubble is outside the constraint of the box. There are additional mechanisms of gas removal including ram-pressure stripping, tidal forces and reionisation, which are not accounted for in our model and are beyond the scope of the paper as we only focus on SN-driven outflows.

Our outflow prescription is simplistic but can be made more physical. A better criterion for the outflow is from \citet{bradamante1998} where they compare the thermal energy from the gas, heated from supernova explosions and stellar winds to the binding energy of the gas (see also \citealt{bertin1992,gibson1994, gibson1997,pipino2004,lanfranchi2004}). We note that the simulation box sizes are larger than the typical half-mass radius of dSphs, but also lower than the virial radius from scaling relations.
 
\begin{figure*}
    \centering
    \includegraphics[width = 1.0\textwidth]{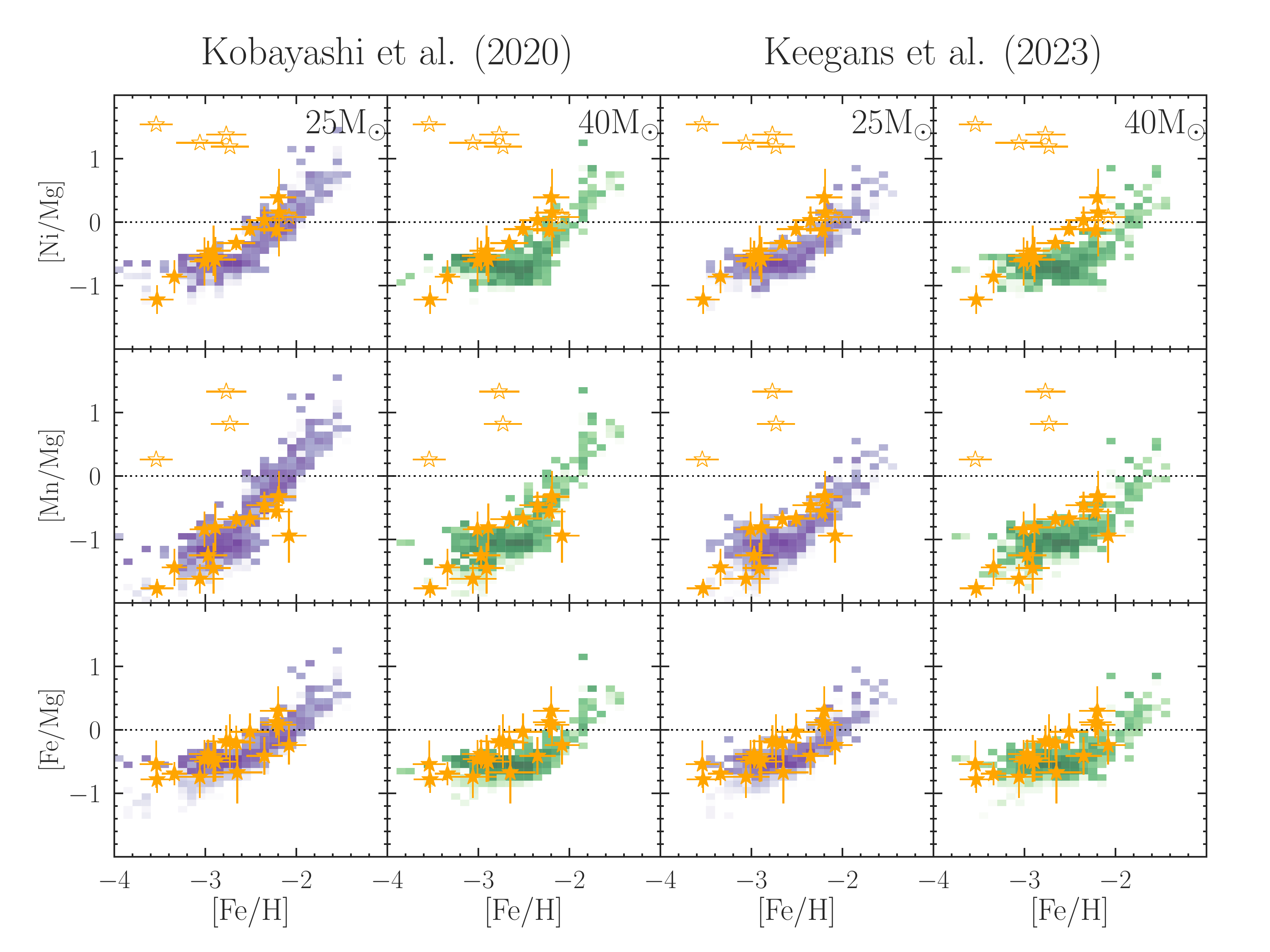}
    \caption{Chemical abundances of our favoured UFD model for [Ni/Mg] (top panels), [Mn/Mg] (middle panels) and [Fe/Mg] (bottom panels) are shown as a function of [Fe/H]. For M$_{\text{ch}}$ \citet{kobayashi2020} (left 2 panels) and \citet{keegans2023} (right 2 panels) SN Ia yields, we change the upper mass limit for CCSNs from $25\,\text{M}_{\sun}$ to $40\,\text{M}_{\sun}$. Model predictions are shown as 2D histograms where each colour map corresponds to the number density of stars within each bin. The orange starred symbols depict the observed chemical abundances in our UFD sample, with empty stars as upper limits.}
    \label{fig:ufd_snia_model}
\end{figure*}

Figure \ref{fig:sfh} shows the model cumulative distribution of the SFH for \textit{dSph A} (top panel) and \textit{dSph B} (bottom panel) in red. We compare them to cumulative SFH in \citet{weisz2014}, who fitted the observed colour-magnitude diagrams from deep Hubble Space Telescope (HST) photometry. The best fit for the SFHs are in black and the total and random uncertainties are grey and yellow shaded areas, respectively. Our \textit{dSph A} toy model has a similar burst of star formation to \citet{weisz2014} at early times, within the total uncertainty boundary for $\leq$ 60\% of the total SFH. However, our model formed the majority of its stars $\approx$ 2 Gyr before \citet{weisz2014}. On the other hand, our \textit{dSph B} toy model has stark differences to \citet{weisz2014}, including their total and random uncertainties. The cumulative SFH from \textit{dSph B} describes a single star-formation episode, with the SFR gradually increasing from the origin until it truncates at $\approx$ 3 Gyr of lookback time. \citet{weisz2014}, however, predicted two star-formation episodes, a first burst followed by a second gradual and extended star-formation episode, separated by a gap of $\approx 2\,\text{Gyr}$. The \textit{dSph B} toy model has its star formation truncated $\approx$ 3 Gyr before \citet{weisz2014}.

\section{Observational Samples}
\label{section:observational}

Our observational samples are taken from a select catalogue of galaxies according to the following criteria. \textit{(i)} The sample must cover a range of metallicities from -4 $\leq$ [Fe/H] $\leq$ -1 to test the full scope of the SNe Ia contributions in our models. \textit{(ii)} The age distribution of the stars in the sample must result from a similar star formation history to the model of comparison. As such, we use these two criteria for our UFD sample from which we select Reticulum II (Ret II, \citealt{ji2016}) and Carina II (Car II, \citealt{ji2020}). The MIKE spectrograph was used by \citet{ji2016} to obtain the spectra of nine Ret II candidates whereas MagLiteS obtained spectra for nine stars in Car II \citep{ji2020}. Both of these UFDs have similar chemical abundances as well as star formation histories, making them prime candidates for our sample.

Our dSph sample consists of Sculptor \citep{north2012, kirby2018, hill2019, delosreyes2020, delosreyes2022}, Leo I \citep{kirby2018, delosreyes2020} and Fornax \citep{kirby2018, delosreyes2020}, with the chemical abundances being taken from a variety of works in the literature, including measurements from the medium-resolution spectroscope DEIMOS \citep{kirby2018, delosreyes2020, delosreyes2022} and the FLAMES/GIRAFFE spectrograph \citep{north2012, hill2019}. There are many differences between the SFH of Sculptor and Leo I. Firstly, \citet{savino2018} found two stellar populations in Sculptor, one which is very old and metal-poor, whose SFH could be fit with a Gaussian centered on $\mu_{1,\tau}=12.58$ Gyr with a standard deviation of $\sigma_{1,\tau}=0.66$ Gyr, and another relatively younger and more metal-rich with $\mu_{2,\tau}=8.0$ and $\sigma_{2,\tau}=0.64$  (see also \citealt{deboer2011}). This is different from \citet{delosreyes2022}, who derived the SFH of Sculptor by using a one-zone GCE model to reproduce the observed chemical abundance patterns, finding that the galaxy formed the vast majority of its stars over a total star formation period of $\sim$ 0.92 Gyr, in agreement with \citet{vincenzo2016} who could reproduce both the observed metallicity distribution function (MDF) and the colour-magnitude diagram (CMD) of the central stellar populations of Sculptor. Even with such differences in the predicted SFH, by following the results of \citet{delosreyes2022}, we deem the chemical abundances of Sculptor to be a good candidate for comparison with dSph A.

Lastly, our sample of Milky Way stars is split into one that covers the iron abundance range [Fe/H] $\leq$ -1 from the Large sky Area Multi-Object fibre Spectroscopic Telescope (LAMOST, \citealt{li2022}) for the Milky Way halo stars, and one that covers the range [Fe/H] $\geq$ -1 from the APOGEE Data Release 17 (DR17) \citep{apogeedr17}. Our MW sample from APOGEE consists of stars residing in both the thin and thick disk, by applying similar cuts as in \citet{weinberg2022}. 

In Figure \ref{fig:snia_obs}, we compare the chemical abundances of [Ni/Mg]-[Fe/H] (top left panel) and [Mn/Mg]-[Fe/H] (bottom left panel) for our chosen observational samples -- UFDs (black stars), Sculptor (red crosses), Leo I (green diamonds) and Fornax (blue squares). Faint blank stars denote upper limits for UFDs. In the same figure, we show several violin plots of our galaxies, depicting their distributions of [Ni/Mg] (top right) and [Mn/Mg] (bottom right). Most works have adopted either the [X/Fe]-[Fe/H] or the [X/Mg]-[Mg/H] diagrams in their work to study the chemical evolution of iron-peak elements in galaxies (e.g. compare \citealt{seitenzahl2013, kirby2019, eitner2020, kobayashi2020, delosreyes2020, delosreyes2022, palla2021, sanders2021, nissen2024}). Throughout our work, we use the [X/Mg]-[Fe/H], as the $x$-axis ([Fe/H]) allows us to calibrate the history of SN Ia enrichment in the model from the convolution of the SFH with the assumed SN Ia DTD, whereas the $y$-axis ([X/Mg]) is an alternative (flipped) version of [$\alpha$/Fe] of the so-called time-delay model (e.g., see \citealt{tinsley1979,matteucci1986}), replacing iron with different iron-peak elements. We find that the distributions of both [Ni/Mg] and [Mn/Mg] increase as we consider galaxies with higher average [Fe/H], with [Mn/Mg] having a stronger dependence than [Ni/Mg]. We fitted a linear regression to calculate the slopes of [Ni/Mg] and [Mn/Mg] as a function of [Fe/H] for our observations within UFDs and dSphs. We find that the [Ni/Mg]-[Fe/H] slope (0.16) is flatter than the [Mn/Mg]-[Fe/H] slope (0.50), indicating that [Mn/Mg] is more metallicity dependent than [Ni/Mg]. Interestingly enough, the steady increase in [Mn/Mg] is in line with the MW halo field stars, suggesting that the main contributors for Mn in both environments took place following a similar process.


\citet{bergemann2008}, \citet{bergemann2019} and \citet{Eitner2023} quantified the impact of non-local thermodynamic equilibrium (NLTE) effects on the Mn and Ni abundances, finding that the chemical abundances of these elements would systematically shift towards higher values at low metallicity when correcting for NLTE effects, with the correction also depending on the stellar atmospheric models that are employed in the abundance analysis (see also \citealt{bergemann2017} and \citealt{amarsi2020} for a comprehensive analysis of Mg and Fe, respectively). This effect may introduce a systematic bias in the observed Ni and Mn, also in the slope of [Ni/Mg] and [Mn/Fe] as a function of [Fe/H]. 
In fact, as the metallicity increases from $\text{[Fe/H]}\approx -2$ to $-1$, the NLTE corrections for both Mn and Ni are found to diminish (e.g., see \citealt{amarsi2020} and \citealt{Eitner2023} for Mn and Ni, respectively), which would shift [Mn/Mg] and [Ni/Mg] more at lower [Fe/H], reducing their slope as a function of [Fe/H].
In our sample only \citet{delosreyes2020} include NLTE corrections to their stars for Mn and Ni in Sculptor, Leo I and Fornax. \citet{kirby2018} only consider the effects of NLTE for Co and Ca, which are elements beyond the scope of this work.


\section{Results}
\label{section:results}

Chemical abundance measurements are the cornerstone of chemical evolution modelling and offer insight into the evolution of galaxies, adding additional dimensions and investigation pathways. This section is split into two parts. In Section \ref{section:ufd} we analyse the predicted chemical abundance patterns from our UFD models with both \citet{kobayashi2020} and \citet{keegans2023} SN Ia yields, including an upper mass limit sensitivity study for the progenitors of CCSNe and a model dispersion comparison between a set number of elements. In Section \ref{section:dsph}, we analyse the predicted chemical abundance patterns for our dSph models, comparing both sub-M$_{\text{ch}}$ and M$_{\text{ch}}$ SN Ia yields from both \citet{kobayashi2020} and \citet{keegans2023}.

\begin{table}

\centering

\resizebox{\columnwidth}{!}{
  
  \begin{tabular}{|c|c|c|c|c|c|}
  
    \hline
    
    \multicolumn{2}{|c|}{} & \multicolumn{2}{|c|}{Kobayashi} & \multicolumn{2}{|c|}{Keegans} \\

    \multicolumn{2}{|c|}{} & 1.0 M$_{\odot}$ & 1.38 M$_{\odot}$ & 1.0 M$_{\odot}$ & 1.4 M$_{\odot}$ \\\hline
    
    \multirow{2}{*}{UFD} & [Ni/Mg] & 0.31 (\textit{0.42}) & 0.36 (\textit{0.50}) & 0.30 (\textit{0.41}) & 0.26 (\textit{0.36}) \\
    & [Mn/Mg] & 0.25 (\textit{0.37}) & 0.31 (\textit{0.41}) & 0.27 (\textit{0.33}) & 0.30 (\textit{0.43}) \\
   
    \hline 

    \multirow{2}{*}{dSph A} & [Ni/Mg] & 0.71 & 0.11 & 0.74 & 0.22 \\
    & [Mn/Mg] & 0.87 \,(\textit{0.83}) & 0.15 \,(\textit{0.14}) & 0.85 \,(\textit{0.79}) & 0.47 \,(\textit{0.46}) \\
   
    \hline 

    \multirow{2}{*}{dSph B} & [Ni/Mg] & 0.61 & 0.25 & 0.72 & 0.16 \\
    & [Mn/Mg] & (\textit{0.90}) & (\textit{0.43}) & (\textit{0.89}) & (\textit{0.34}) \\
   
    \hline 
  
  \end{tabular}}
  
  \caption{For all three toy models, elemental abundances and progenitors, we compute the statistics from the KS-test to find the likelihood that our models and their respective observations are drawn from the same distribution. In the table, we report the maximum vertical difference between the predicted and observed cumulative distributions. Higher values correspond to larger distances between the two distributions and thus poorer KS-test statistics. Values for the UFD model are for the 40 M$_{\odot}$ upper-mass limit, whereas the italic values within parenthesis are for the 25 M$_{\odot}$ limit. For \textit{dSph A} and \textit{dSph B}, both values are for the 25 M$_{\odot}$ upper-mass limit, however, the italic values within parenthesis report the KS-test statistics when using only chemical abundance measurements with NLTE corrections.}
  
  \label{table:statistic}
\end{table}

\subsection{Ultra-Faint Dwarf Galaxies}
\label{section:ufd}

\begin{figure*}
    \centering
    \includegraphics[width = 1.00\textwidth]{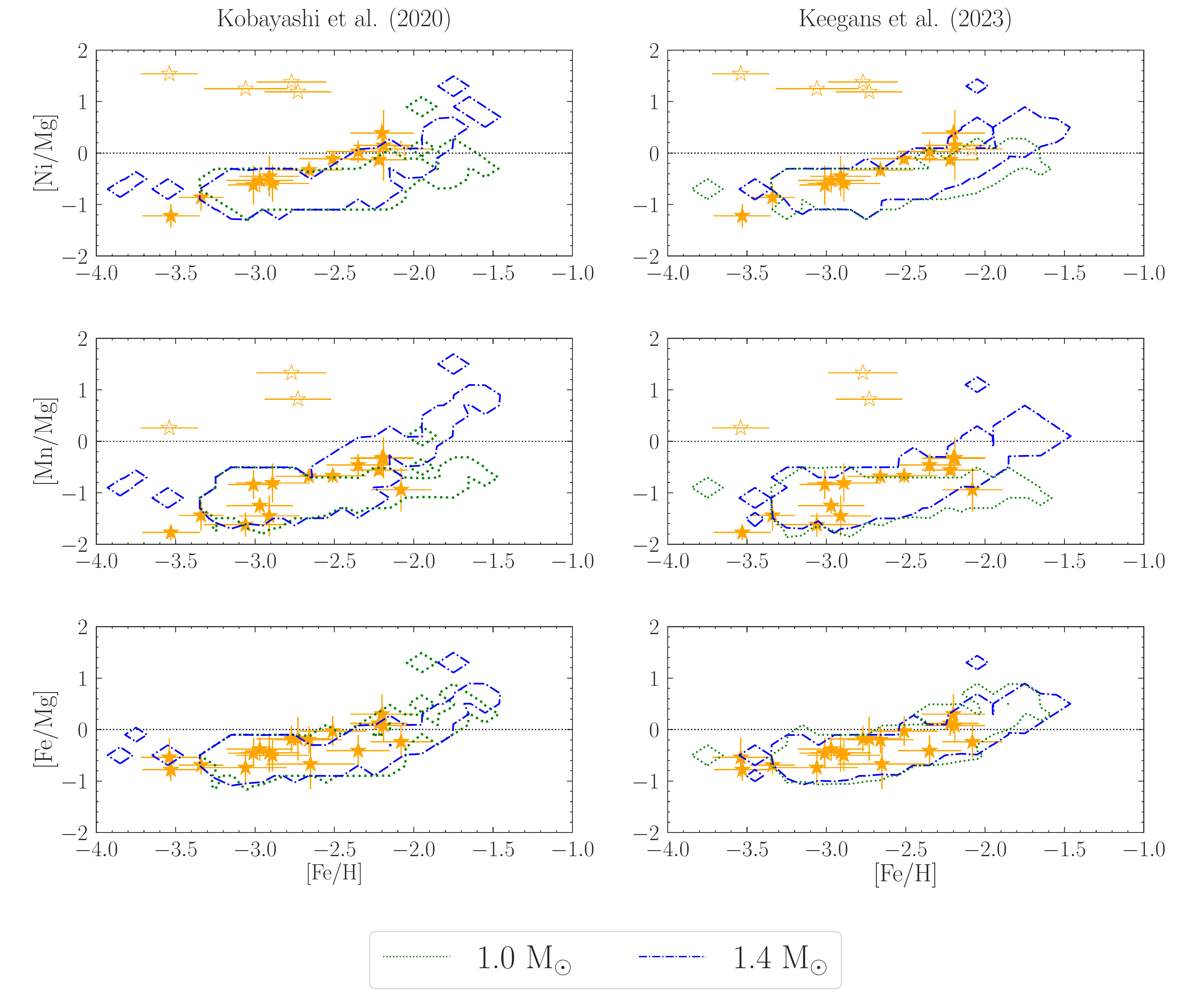}
    \caption{The [Ni/Mg]-[Fe/H] (top panels), [Mn/Mg]-[Fe/H] (middle panels) and [Fe/Mg]-[Fe/H] (bottom panels) chemical abundances of our UFD toy models assuming the \citet{kobayashi2020} (left column) and \citet{keegans2023} (right column) SN Ia yields. The results of the models assuming SN Ia progenitors with masses $M_{\text{WD}}=1.0\,\text{M}_{\sun}$ are the green dotted contours, and $M_{\text{WD}}\approx1.4\,\text{M}_{\sun}$ are the blue dot-dashed contours corresponding to densities of 99.9 per cent normalised to the maximum value of each distribution. The chemical abundances of our UFD samples are the orange starred symbols, with empty stars as upper limits.}
    \label{fig:ufd_mixture}
\end{figure*}

We begin our analysis by exploring the combined effect of different upper mass limits for CCSNe and M$_{\text{ch}}$ SN Ia yields. In particular, Fig. \ref{fig:ufd_snia_model} shows a comparison between different [X/Mg]-[Fe/H] chemical abundance patterns in our UFD models, where X corresponds to Ni (top panels), Mn (middle panels) and Fe (bottom panels). The first two columns show the predicted chemical abundances as purple and green 2D histograms for the 1.38 M$_{\text{ch}}$ SN Ia yields of \citet{kobayashi2020}, where we vary the upper mass limit for CCSNe ($M_{\text{up}}=25\,\text{M}_{\sun}$ in purple and $40\,\text{M}_{\sun}$ in green). In short, any star which is above the shown mass will not explode as a CCSN. The third and fourth columns show the same abundance patterns for a different massive-star cutoff as the first two, but for the \citet{keegans2023} M$_{\text{ch}}$ SN Ia yields. Orange stars denote the observed chemical abundances of our UFD sample. As UFDs have significantly fewer stars, and thus supernovae events, than dSphs, the upper mass limit for CCSN is expected to have more of an impact on them.

Both \citet{kobayashi2020} and \citet{keegans2023} M$_{\text{ch}}$ yields predict similar chemical abundance patterns of [Mn/Mg]-[Fe/H] and [Ni/Mg]-[Fe/H], which reproduce the observed sample of RGB abundances in UFD galaxies when assuming a $25\,\text{M}_{\sun}$ upper mass limit for CCSNe. A $40\,\text{M}_{\sun}$ upper mass limit produces more Fe in massive stars, allowing [Fe/H] to evolve further before SN Ia events start dumping Fe-peak elements into the ISM, increasing [X/Mg]. Note that the total amount of Fe released by a massive star population is higher by $\approx 1.5$ orders of magnitude than Ni and  $\approx 2.5$ orders than Mn (see Fig. \ref{fig:snia_ccsn_comp}), hence the main effect of increasing the upper mass limit is shifting the predicted chemical abundance patterns to the right, keeping the same characteristic shape. 
 
Models with the \citet{kobayashi2020} and \citet{keegans2023} SN Ia yields predict similar [Fe/Mg]-[Fe/H] chemical abundance patterns for both 25 and $40\,\text{M}_{\sun}$ upper mass limits. Our toy model however can reach higher super-solar [Mn/Mg] with the \citet{kobayashi2020} SN Ia yields than \citet{keegans2023}, indicative of the differences between their respective Mn production. Interestingly, there have been no CCSN progenitor stars with initial mass $M \geq 30 \, \text{M}_{\sun}$ which have been observed \citep{smartt2009, kobayashi2020b}. For the continuation of our work, a $25\,\text{M}_{\sun}$ upper mass limit for CCSNe is adopted in our reference toy models, following also previous chemical evolution studies (see \citealt{prantzos2018,vincenzo2018a,vincenzo2018b,kobayashi2020b} and references therein).

\begin{figure*}
    \centering
    \includegraphics[width = 1.0\textwidth]{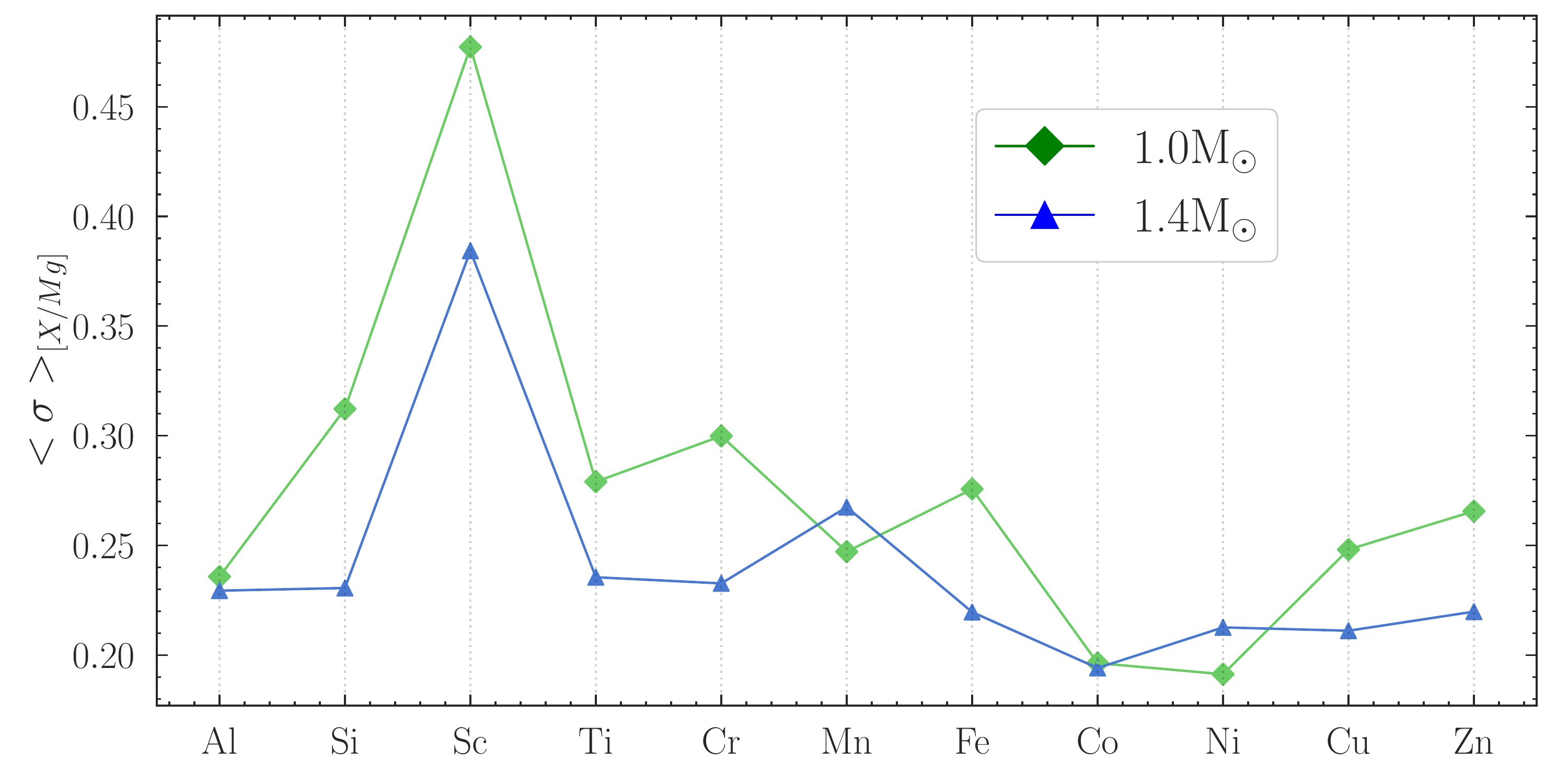}
    \caption{Distribution of the average standard deviation of [X/Mg] from our UFD model where green triangles and blue triangles correspond to the predictions of the models with \citet{keegans2023} SN Ia progenitors with masses $M_{\text{WD}}=1.0\,\text{M}_{\sun}$ and $1.4\,\text{M}_{\sun}$, respectively.
    }
    \label{fig:ufd_dispersion} 
\end{figure*}

Having established the role of different upper mass limits for CCNSe, we not turn our attention in characterising the behaviour of chemical evolution models assuming SN Ia progenitors with different masses, comparing the predictions of different works available in the literature. Fig. \ref{fig:ufd_mixture} compares the predicted UFD chemical abundance patterns that are obtained when assuming the \citet[first column]{kobayashi2020} and \citet[second column]{keegans2023} SN Ia progenitors. Chemical evolution models with the $1.0\,\text{M}_{\sun}$ sub-M$_{\text{ch}}$ and $1.4\,\text{M}_{\sun}$ M$_{\text{ch}}$ models are shown through blue and contours, respectively. Filled orange stars are the observed chemical abundances in our sample of UFD stars. Both sub-M$_{\text{ch}}$ and M$_{\text{ch}}$ SN Ia yields of \citet{kobayashi2020} predict similar abundance patterns to observations at very low metallicity (i.e. [Fe/H] $\lesssim$ -2) for all given elements; however, the UFD model with M$_{\text{ch}}$ SN Ia progenitors predict a larger production of Fe-peak elements to the ISM than those with sub-M$_{\text{ch}}$ progenitors, reproducing the observed patterns also at higher metallicities. We also conduct a two-sample Kolmogorov-Smirnov test if the chemical abundances from our observations and model can be drawn from identical distributions. For each chemical element, we compared the normalised cumulative distribution of the model and observed stars and found the maximum vertical difference between the two, which we dub the \enquote{statistic}. Table \ref{table:statistic} displays the statistics between the model and observations, where lower values indicate a higher probability that the two samples are drawn from the same distributions.
Our results in Table \ref{table:statistic} confirm that both SN Ia progenitors have similar statistics for [Ni/Mg] and [Mn/Mg] for our UFD toy model, with slight differences between SN Ia yields. \citet{kobayashi2020} sub-M$_{\text{ch}}$ and \citet{keegans2023} M$_{\text{ch}}$ 
SN Ia progenitors provide the best agreement to observational data. 

The $1.4\,\text{M}_{\sun}$ M$_{\text{ch}}$ SN Ia progenitors of \citet{keegans2023} are found to exhibit similar [Ni/Mg]-[Fe/H] and [Mn/Mg]-[Fe/H] to those from \citet{kobayashi2020}, reproducing the abundance patterns seen within our UFD sample. As previously stated, higher [Mn/Mg] is reached with the \citet{kobayashi2020} M$_{\text{ch}}$ SN Ia yields than \citet{keegans2023}, as Mn is produced more in the former (see Fig \ref{fig:yields_comp}). Interestingly, there are negligible differences in [Fe/Mg]-[Fe/H] abundance ratios between \citet{kobayashi2020} and \citet{keegans2023} sub-M$_{\text{ch}}$ and M$_{\text{ch}}$ SN Ia yields. 

Fig. \ref{fig:ufd_dispersion} shows the average standard deviation of the chemical abundances as predicted by our UFD model assuming the \citet{keegans2023} SN Ia yields. We remind the readers that a $25\,\text{M}_{\sun}$ massive star cut-off for CCSN progenitors is assumed in the reference UFD model. We show more elements than the usual four to highlight the disparity between them for two SNe Ia progenitor masses: $1.0\,\text{M}_{\sun}$ (green diamonds) and $1.4\,\text{M}_{\sun}$ (blue triangles). The average model dispersion is estimated through the arithmetic mean of the standard deviation of [X/Mg] through numerous bins of [Fe/H]. Elements on the $x$-axis are in ascending order of atomic number. We note that the dispersions in Fig. \ref{fig:ufd_dispersion} are from one model, and a more accurate estimate could be obtained by averaging out the dispersion over several runs of the same model.


With the exception of Sc, the M$_{\text{ch}}$ model in Fig. \ref{fig:ufd_dispersion} has an almost constant average dispersion that ranges between $0.20$ and $0.25$ for the considered chemical elements while clearer differences emerge between progenitor masses. An interesting feature of Fig. \ref{fig:ufd_dispersion} is the noticeably higher average dispersion of [Sc/Mg] than other abundance ratios, with the sub-M$_{\text{ch}}$ SNe Ia progenitor model predicting a higher dispersion than the M$_{\text{ch}}$ model. Although \citet{hill2019} measured weak Sc lines, they found an unusually large dispersion of [Sc/Mg] at [Fe/H] $\approx$ -2.3 relative to other elements they analysed. Abundance ratios like [Sc/Mg] with large predicted dispersion may be used to pinpoint specific enrichment events in the model, determining also whether there is any systematic (anti)correlation with the residuals of other chemical elements that could be coupled to identify those events with more precision, however, we note that the uncertainty in Sc abundance measurements may be as high as $\approx 0.1\,\text{dex}$. The study of how the correlation coefficients of the residuals of different elements change as a function of metallicity in relation to the chemical evolution of galaxies will be one of the subjects of our future work.





\subsection{Dwarf Spheroidal Galaxies}
\label{section:dsph}

\begin{figure*}
    \centering
    \includegraphics[width = 1.0\textwidth]{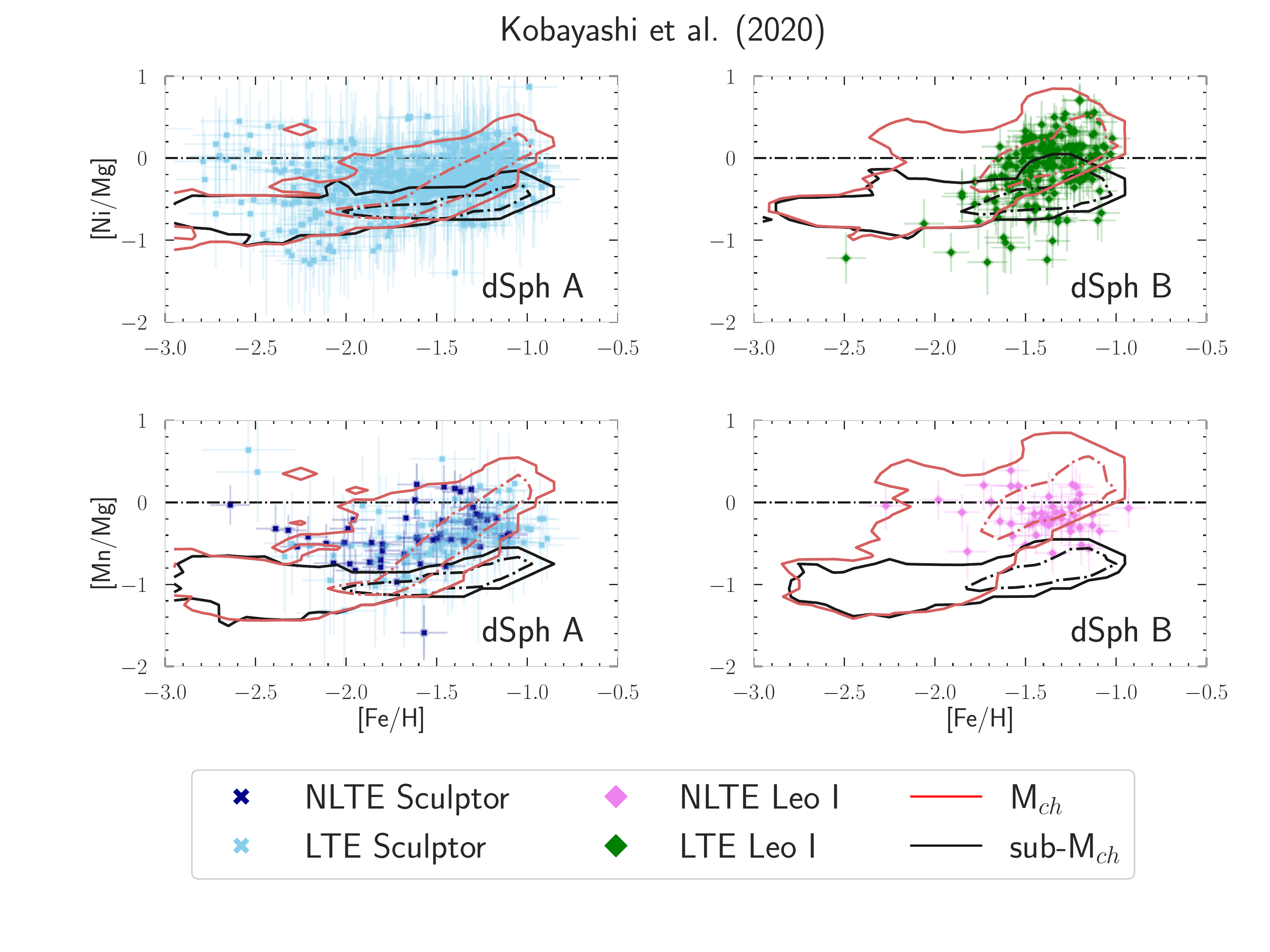}
    \caption{The chemical abundance patterns of [Ni/Mg]-[Fe/H] (top panels) and [Mn/Mg]-[Fe/H] (bottom panels) for our two toy dSph models, where \textit{dSph A} (left column) simulates the SFH and chemical enrichment of an ancient dSph galaxy like Sculptor, and \text{dSph B} (right column) is for a dSph galaxy with a more extended period of star formation like Leo I. The predictions of the dSph models assuming the 1.38 M$_{\text{ch}}$ and 1.0 sub-M$_{\text{ch}}$ SN Ia yields of \citet{kobayashi2020} are shown as red and black contours, respectively, corresponding to densities of 99.99 per cent (thick solid lines) and 90 per cent (thin dotted lines) normalised to the maximum value for each distribution. Both models adopt a stellar mass cut-off of $25\,\text{M}_{\sun}$ for CCSNe. Model predictions are compared to the observed samples in Sculptor (light blue and dark blue crosses for LTE and NLTE, respectively) and Leo I (green and pink diamonds for LTE and NLTE, respectively).}
    \label{fig:dsph_abund_kob}
\end{figure*}

\begin{figure*}
    \centering
    \includegraphics[width = 1.0\textwidth]{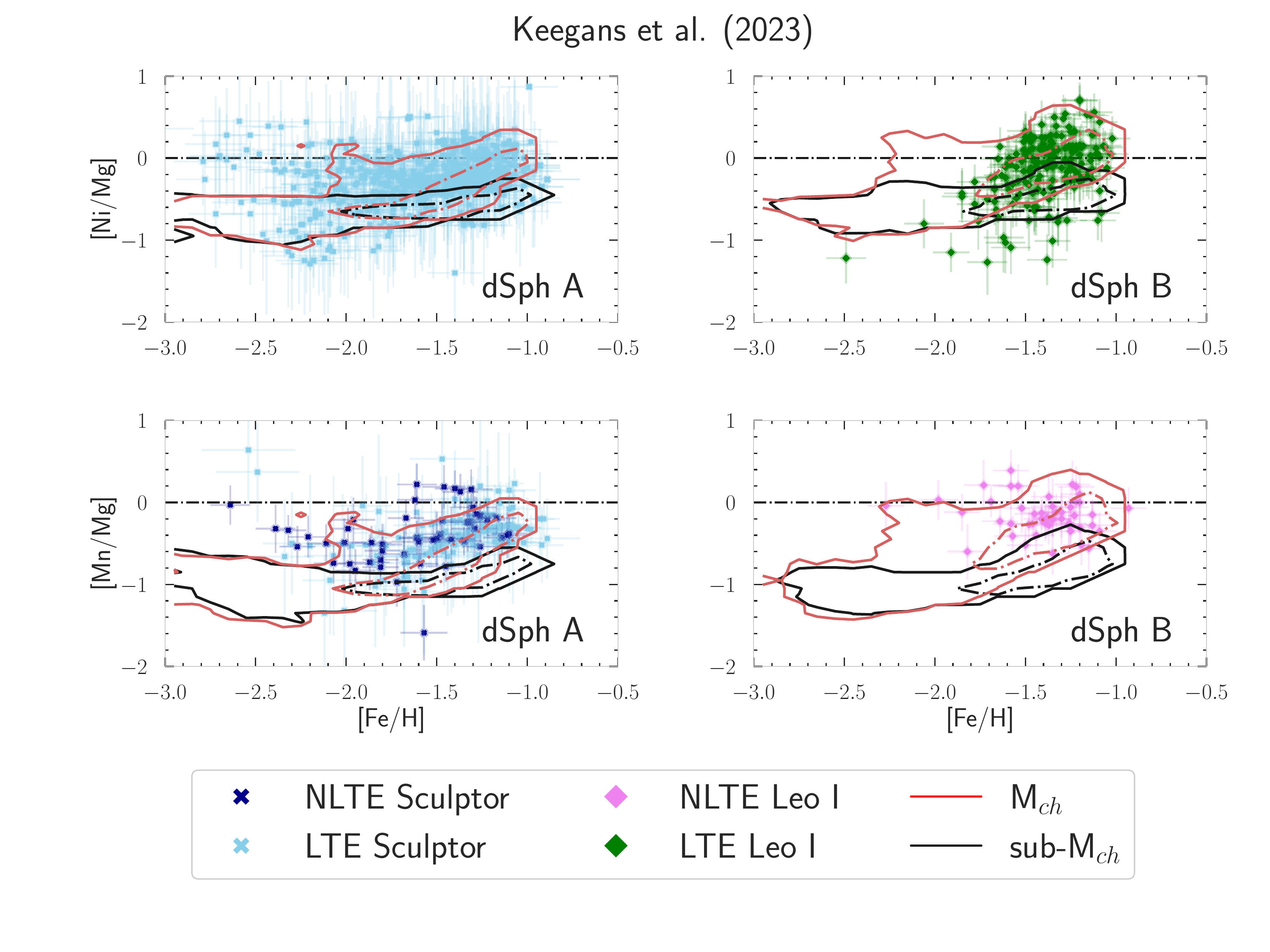}
    \caption{Similar to Figure \ref{fig:dsph_abund_kob}, but for 1.4 M$_{\text{ch}}$ and 1.0 sub-M$_{\text{ch}}$ \citet{keegans2023} SN Ia yields.}
    \label{fig:dsph_abund}
\end{figure*}

Fig. \ref{fig:dsph_abund_kob} shows a comparison between the predictions of our toy models for dSph galaxies assuming the \citet{kobayashi2020} SN Ia yields and the observational data of [Ni/Mg]-[Fe/H] (top panels) and [Mn/Mg]-[Fe/H] (bottom panels), considering dSph galaxies with different SFHs (different columns); the \textit{dSph A} model on the left characterises an ancient dSph galaxy like Sculptor, whereas \textit{dSph B} on the right is for a dSph galaxy with a more extended star formation activity like Leo I. The dSph models with the \citet{kobayashi2020} $1.38\,\text{M}_{\sun}$ M$_{\text{ch}}$ and $1.0\,\text{M}_{\sun}$ \text{sub}-M$_{\text{ch}}$ SN Ia yields are presented as red and black contours, respectively. Our model stars are compared to the chemical abundances of Sculptor (\textit{dSph A}) and Leo I (\textit{dSph B}), which are shown as blue crosses and green diamonds, respectively. 

The \textit{dSph A} model with the \citet{kobayashi2020} $1.38\,\text{M}_{\sun}$ SN Ia progenitors can reproduce the observed abundance patterns of [Ni/Mg]-[Fe/H] and [Mn/Mg]-[Fe/H] in Sculptor, being also characterized by the lowest KS-test statistics (see second row in Table \ref{table:statistic}). Sculptor stars are found to have a large dispersion in [Ni/Mg] and [Mn/Mg] at [Fe/H] $\lesssim$ -2, and our \textit{dSph A} model reproduces some of the abundances at these very low metallicities, including several stars with sub-solar [Ni/Mg] and [Mn/Mg]. \textit{dSph A} also predicts a second, less dense sequence of stars with super-solar [Ni/Mg] and [Mn/Mg]. The \textit{dSph B} model with the \citet{kobayashi2020} $1.38\,\text{M}_{\sun}$ SN Ia progenitors predicts that stars have a larger scatter in their abundances at [Fe/H] $\lesssim$ -1.6 than the \textit{dSph A} model, with a more pronounced second sequence of stars with super-solar [Ni/Mg] and [Mn/Mg]. \textit{dSph B} also predicts a sharper increase of [Ni/Mg] and [Mn/Mg] from $\text{[Fe/H]}\approx -1.6$, reproducing the super-solar [Ni/Mg] and [Mn/Mg] at those higher metallicities that are observed in Leo I. 

The \textit{dSph A} model with the \citet{kobayashi2020} $1.0\,\text{M}_{\sun}$ SN Ia progenitors cannot reproduce the overall chemical abundance patterns observed in Sculptor, though it can explain the observed [Ni/Mg] of several stars with iron abundances in the range $-1.5 \lesssim \text{[Fe/H]}\lesssim -0.8$ that have [Ni/Mg] lower than the predictions of the \textit{dSph A} model with $1.4\,\text{M}_{\sun}$ M$_{\text{ch}}$ SN Ia progenitors. In the observed sample, there is also a clump of stars with $ \text{[Mn/Mg]} \approx -0.5$ at iron abundances $-1.2 \lesssim \text{[Fe/H]}\lesssim -0.8$, corresponding to intermediate values of [Mn/Mg] between those predicted by the $1\,\text{M}_{\sun}$ sub-M$_{\text{ch}}$ SN Ia progenitors and the $1.4\,\text{M}_{\sun}$ M$_{\text{ch}}$ ones, \textit{(i)} showing that the Mn yields from SNe Ia of \citet{kobayashi2020} are more sensitive to the SN Ia progenitor mass than the Ni yields (see also Fig. \ref{fig:yields_comp}), and \textit{(ii)} suggesting the presence of some contribution from a population of sub-M$_{\text{ch}}$ SN Ia progenitors that have masses larger than $1\,\text{M}_{\sun}$ explored in this work. Similar arguments would also be valid from the analysis of the behaviour of \textit{dSph B} with the \citet{kobayashi2020} $1\,\text{M}_{\sun}$ SN Ia progenitors, though the observed sample shows a larger scatter in [Ni/Mg], with a dozen of stars that have lower [Ni/Mg] than both models in the figure. 

Fig. \ref{fig:dsph_abund} is similar to Fig. \ref{fig:dsph_abund_kob} but shows the predictions of our dSph models with the \citet{keegans2023} SN Ia yields. Similarly to Fig. \ref{fig:dsph_abund_kob}, the chemical abundances of our observations are shown as red crosses and green diamonds for Sculptor and Leo I, respectively. Both \textit{dSph A} and \textit{dSph B} models with only $1.4\,\text{M}_{\sun}$ M$_{\text{ch}}$ SN Ia progenitors reproduce better the observed [Ni/Mg]-[Fe/H] and [Mn/Mg]-[Fe/H] chemical abundance patterns, showing a similar average behaviour as the models with the \citet{kobayashi2020} SN Ia yields but predicting \textit{(i)} lower values of [Ni/Mg] and [Mn/Mg] at $-1.5 \lesssim \text{[Fe/H]}\lesssim -1.0$ and \textit{(ii)} a less sharp increase of [Mn/Mg] in \textit{dSph B} from $\text{[Fe/H]}\approx -1.6$ because of the lower amount of Mn from the M$_{\text{ch}}$ SN Ia models of \citet{keegans2023} than those of \citet{kobayashi2020}; this allows \textit{dSph B} with \citet{keegans2023} SN Ia yields to reproduce better the observed [Mn/Mg] in Leo I -- the same SN Ia yields, however, do not allow \textit{dSph A} to reach the high [Mn/Mg] that are also observed in Sculptor, which are better reproduced by the \citet{kobayashi2020} M$_{\text{ch}}$ SN Ia yields. Our novel outflow prescription as well as our chemical evolution model serve as key differences between our results and those from literature (see \citealt{kobayashi2020, delosreyes2020, delosreyes2022}). With identical yields to \citet{kobayashi2020}, we find M$_{\text{ch}}$ SN Ia progenitors can reproduce the chemical abundance patterns in dSphs, whereas \citet{kobayashi2020} found sub-M$_{\text{ch}}$ SN Ia progenitors to be the most dominant. Future theoretical work on SN Ia yields and more observations on dSphs may fix some of these discrepancies. 

There are several outliers scattered throughout observations for both [Ni/Mg]-[Fe/H] and [Mn/Mg]-[Fe/H] at both low and high metallicities. Sculptor [Mn/Mg]-[Fe/H] includes outliers from LTE and NLTE observations, which \textit{dSph A} cannot reproduce, either with M$_{\text{ch}}$ or sub-M$_{\text{ch}}$ SN Ia progenitors. The CMD-fitting analysis of \citet{weisz2014} from deep HST photometry determined smoother SFHs than those predicted by our \textit{dSph A} and \textit{dSph B} models for Sculptor and Leo I, respectively (see Fig. \ref{fig:sfh} and Section \ref{section:models}), which would cause a more gentle build-up of the stellar mass and thus less numbers of prompt SNe Ia, flattening the slope of [X/Mg]-[Fe/H] and shifting them to the left, towards lower [Fe/H] values.

Interestingly enough, we find that our chemical evolution models with sub-M$_{\text{ch}}$ SN Ia progenitors return much higher (poorer) KS-test statistics than M$_{\text{ch}}$ SN Ia progenitors, regardless of the SN Ia yield prescription. For \textit{dSph A}, \citet{kobayashi2020} SN Ia yields provide lower (better) KS-test statistics for M$_{\text{ch}}$ SN Ia progenitors than \citet{keegans2023}, with the opposite being true for the \textit{dSph B} toy model. We note that applying NLTE corrections to our observed sample of uncorrected Mn and Ni abundances would systematically shift them towards higher values, further increasing their distance from the sub-M$_{\text{ch}}$ SN Ia models.

\section{Conclusions}
\label{section:conclusions}

In our present work, we investigate the \citet{kobayashi2020} and \citet{keegans2023} metallicity-dependent SN Ia yields on the chemical evolution of UFDs and dSphs, exploring also the effect of M$_{\text{ch}}$ and sub-M$_{\text{ch}}$ SN Ia progenitors. Our analysis is focused on Mn and Ni, as the yields of these elements show the largest changes as a function of the WD progenitor mass (see Section \ref{section:type1a_yields} for a discussion on these SN Ia yields). We incorporate these sets of SN Ia yields within our inhomogeneous chemical evolution tool {\tt{i-GEtool}} \citep{alexander2023} in Section \ref{section:models} and create one UFD and two dSph models, assuming the massive star yields of \citet{kobayashi2006} and the AGB yields of \citet{karakas2010}, which are both kept constant as we change the SNe Ia yields. The assumed DTD for SNe Ia is $\propto \tau^{-1.1}$ as a function of the delay-time, $\tau$, which does not depend on metallicity and is same for all progenitors (see, e.g., \citealt{kobayashi2020} and \citealt{johnson2023} for detailed discussions on this). Each model assumes the galaxy to form from a single accretion event which produces SFHs that truncate at different cosmic times. The IMF is from \citet{kroupa2001}, and the outflow prescriptions are the same as \citet{alexander2023}. In Section \ref{section:results}, we compare our observational sample of UFD stellar abundances (Carina II from \citealt{ji2020} and Reticulum II from \citealt{ji2016}) to the respective model, first examining the effect of changing the massive star cut-off and then the various SN Ia progenitor yields. We also compare the average chemical abundance dispersion of several chemical elements relative to Mg, using the \citet{keegans2023} M$_{\text{ch}}$ SN Ia yields. Finally, we examine the predicted [Ni/Mg]-[Fe/H] and [Mn/Mg]-[Fe/H] from our two dSph models and compare them to observations in Sculptor and Leo I, considering both M$_{\text{ch}}$ and sub-M$_{\text{ch}}$ SNe Ia progenitors. The following is a summary of the conclusions in this present work.

\begin{itemize}

    \item[(i)] M$_{\text{ch}}$ SN Ia progenitors contribute more Fe-peak elements to the ISM than sub-M$_{\text{ch}}$. The SN Ia yields of Mn show a stronger dependence on the mass of the WD progenitor than Ni. The $0.8\,\text{M}_{\sun}$ sub-M$_{\text{ch}}$ SN Ia model contributes the least Ni, Mn and Fe yields, but interestingly the most for Mg. For metallicities $\log(Z / Z_{\sun}) \lesssim -2$, the W7 metallicity-independent SN Ia yields of \citet{iwamoto1999} are similar to the other M$_{\text{ch}}$ SN Ia yields that are examined in this work. 
    
    \item[(ii)] A single SN Ia event provides comparable amounts of iron-peak elements to the ISM as that of all massive stars forming out of a $10^{4}\,\text{M}_{\sun}$ stellar population, which makes the stellar populations in the local UFD and dSph galaxies probably the best laboratories in the cosmos to study the SN Ia nucleosynthesis. Large differences in the IMF-averaged massive-star yields are found when changing the cutoff mass between 25 and 40 M$_{\sun}$ at metallicities $\log(Z/Z_{\sun}) \lesssim -1$. Such difference decreases as $Z$ approaches $Z_{\sun}$.

    \item[(iii)] We compare observational samples taken from different galaxy environments, spanning more than 10,000 individual stars, focusing on the observed [Ni/Mg]-[Fe/H] and [Mn/Mg]-[Fe/H] abundance patterns. [Mn/Mg] evolves more strongly than [Ni/Mg] as a function of [Fe/H], showing the same average behaviour across the different galaxy environments, suggesting that the origin of Ni and Mn in both the MW field stars and dSphs may be caused by similar producers. Applying NLTE corrections to our observed sample of uncorrected Mn and Ni abundances would systematically shift them towards higher values (e.g., see \citealt{amarsi2020} and \citealt{Eitner2023} for Mn and Ni, respectively). The fact that NLTE corrections for both Mn and Ni diminish when moving from $\text{[Fe/H]}\approx -2$ to $-1$ would cause the shift towards higher values to be stronger at lower [Fe/H], reducing the slope of the [Mn/Mg]-[Fe/H] and [Ni/Mg]-[Fe/H] abundance patterns.

    \item[(iv)] In our toy model for UFD galaxies, a $25\,\text{M}_{\sun}$ cut-off reproduces the observations of Carina II and Reticulum II while also maintaining a consistent SFH similar to the observed one, as higher upper mass limits truncate the SFH quicker, stopping any further chemical enrichment. 
    
    \item[(v)] The assumption of only M$_{\text{ch}}$ SNe Ia progenitors produces similar to identical chemical abundance patterns to those observed, with the 1.0 sub-M$_{\text{ch}}$ SN Ia models under producing [Ni/Mg] and [Mn/Mg] for both dSph models. M$_{\text{ch}}$ and sub-M$_{\text{ch}}$ produce similar abundance patterns in our UFD model, with differences between the abundances dependent on the adopted SN Ia yields.

    \item[(vi)] We examined the predicted average dispersion in the chemical abundances of our model UFD stars. The dispersion of [Sc/Mg] exhibits the highest values among the iron-peak elements relative to Mg, similar to the findings of \citet{hill2019}. Of the three SNe Ia progenitor models that are explored in this work, the $M_{\text{ch}}$ model has an almost constant dispersion when considering different chemical elements, excluding Sc. 

    \item[(vii)] Our toy model \textit{dSph A} has a short but powerful SFH like in the Sculptor dSph galaxy (e.g., \citealt{deboer2011}) whereas \textit{dSph B} is characterised by a long but calm SFH, similar to the observations in Leo I (e.g., \citealt{weisz2014}). When assuming only M$_{\text{ch}}$ SN Ia progenitors, both \textit{dSph A} and \textit{dSph B} provide a good match to the observed [Mn/Mg]-[Fe/H] and [Ni/Mg]-[Fe/H]. \citet{kobayashi2020} M$_{\text{ch}}$ SN Ia progenitors produce more Mn than \citet{keegans2023} M$_{\text{ch}}$ progenitors, resulting with higher [Mn/Mg] at $\text{[Fe/H]} \gtrsim -1$. 
    Models with only sub-M$_{\text{ch}}$ SN Ia progenitors systematically under produce both Mn and Ni in \textit{dSph A} and \textit{dSph B}.



   \item[(viii)] Our \textit{dSph A} and \textit{dSph B} toy models assuming the \citet{kobayashi2020} SN Ia yields reproduce more outliers in the chemical abundance patterns than \citet{keegans2023} at both low and high metallicities.

\end{itemize}

Our present work shows that M$_{\text{ch}}$ SN Ia progenitors in inhomogeneous chemical evolution models can reproduce the [Mn/Mg]-[Fe/H] and [Ni/Mg]-[Fe/H] as observed in a sample of red giants in local UFD and dSphs, without the need of a substantial fraction of sub-M$_{\text{ch}}$ SN Ia progenitors. Our results indicate the importance of accounting for inhomogeneous chemical enrichment and metallicity-dependent SN Ia yields, which are the main aspects distinguishing our work from the previous chemical evolution studies of iron-peak elements. In future work, we hope to also include neutron capture elements in our inhomogeneous chemical evolution model to further study Ret II as it is believed to be a rich source of $r$-process elements \citet{ji2016}. Finally, our models can be applied to the Solar neighbourhood to examine the chemical evolution of several elements and probe the star formation in the nearby volume of our Galaxy in great detail. 

\section*{Acknowledgements}

We thank an anonymous referee for their comments and feedback which greatly improved the quality of our work. We would also like to thank both James Keegans and Chiaki Kobayashi for valuable discussions on SN Ia as well as their yields for various SN Ia progenitors.

\section*{Data Availability}

The data within this work, which includes the chemical abundances as predicted by our inhomogeneous galactic chemical evolution models for UFDs and dSphs, will be shared upon reasonable request to the lead author.


\bibliographystyle{mnras}
\bibliography{SNIa} 

\begin{thebibliography}{}
\makeatletter
\relax
\def\mn@urlcharsother{\let\do\@makeother \do\$\do\&\do\#\do\^\do\_\do\%\do\~}
\def\mn@doi{\begingroup\mn@urlcharsother \@ifnextchar [ {\mn@doi@}
  {\mn@doi@[]}}
\def\mn@doi@[#1]#2{\def\@tempa{#1}\ifx\@tempa\@empty \href
  {http://dx.doi.org/#2} {doi:#2}\else \href {http://dx.doi.org/#2} {#1}\fi
  \endgroup}
\def\mn@eprint#1#2{\mn@eprint@#1:#2::\@nil}
\def\mn@eprint@arXiv#1{\href {http://arxiv.org/abs/#1} {{\tt arXiv:#1}}}
\def\mn@eprint@dblp#1{\href {http://dblp.uni-trier.de/rec/bibtex/#1.xml}
  {dblp:#1}}
\def\mn@eprint@#1:#2:#3:#4\@nil{\def\@tempa {#1}\def\@tempb {#2}\def\@tempc
  {#3}\ifx \@tempc \@empty \let \@tempc \@tempb \let \@tempb \@tempa \fi \ifx
  \@tempb \@empty \def\@tempb {arXiv}\fi \@ifundefined
  {mn@eprint@\@tempb}{\@tempb:\@tempc}{\expandafter \expandafter \csname
  mn@eprint@\@tempb\endcsname \expandafter{\@tempc}}}

\bibitem[\protect\citeauthoryear{{Abdurro'uf} et~al.,}{{Abdurro'uf}
  et~al.}{2022}]{apogeedr17}
{Abdurro'uf} et~al., 2022, \mn@doi [\apjs] {10.3847/1538-4365/ac4414}, \href
  {https://ui.adsabs.harvard.edu/abs/2022ApJS..259...35A} {259, 35}

\bibitem[\protect\citeauthoryear{{Alexander}, {Vincenzo}, {Ji}, {Richstein},
  {Jordan}  \& {Gibson}}{{Alexander} et~al.}{2023}]{alexander2023}
{Alexander} R.~K.,  {Vincenzo} F.,  {Ji} A.~P.,  {Richstein} H.,  {Jordan}
  C.~J.,   {Gibson} B.~K.,  2023, \mn@doi [\mnras] {10.1093/mnras/stad1312},
  \href {https://ui.adsabs.harvard.edu/abs/2023MNRAS.522.5415A} {522, 5415}

\bibitem[\protect\citeauthoryear{{Amarsi} et~al.,}{{Amarsi}
  et~al.}{2020}]{amarsi2020}
{Amarsi} A.~M.,  et~al., 2020, \mn@doi [\aap] {10.1051/0004-6361/202038650},
  \href {https://ui.adsabs.harvard.edu/abs/2020A&A...642A..62A} {642, A62}

\bibitem[\protect\citeauthoryear{{Andrews}, {Weinberg}, {Sch{\"o}nrich}  \&
  {Johnson}}{{Andrews} et~al.}{2017}]{andrews2017}
{Andrews} B.~H.,  {Weinberg} D.~H.,  {Sch{\"o}nrich} R.,   {Johnson} J.~A.,
  2017, \mn@doi [\apj] {10.3847/1538-4357/835/2/224}, \href
  {https://ui.adsabs.harvard.edu/abs/2017ApJ...835..224A} {835, 224}

\bibitem[\protect\citeauthoryear{{Arnett}}{{Arnett}}{1996}]{arnett1996}
{Arnett} D.,  1996, {Supernovae and Nucleosynthesis: An Investigation of the
  History of Matter from the Big Bang to the Present}

\bibitem[\protect\citeauthoryear{{Bergemann} \& {Gehren}}{{Bergemann} \&
  {Gehren}}{2008}]{bergemann2008}
{Bergemann} M.,  {Gehren} T.,  2008, \mn@doi [\aap]
  {10.1051/0004-6361:200810098}, \href
  {https://ui.adsabs.harvard.edu/abs/2008A&A...492..823B} {492, 823}

\bibitem[\protect\citeauthoryear{{Bergemann}, {Collet}, {Amarsi}, {Kovalev},
  {Ruchti}  \& {Magic}}{{Bergemann} et~al.}{2017}]{bergemann2017}
{Bergemann} M.,  {Collet} R.,  {Amarsi} A.~M.,  {Kovalev} M.,  {Ruchti} G.,
  {Magic} Z.,  2017, \mn@doi [\apj] {10.3847/1538-4357/aa88cb}, \href
  {https://ui.adsabs.harvard.edu/abs/2017ApJ...847...15B} {847, 15}

\bibitem[\protect\citeauthoryear{{Bergemann} et~al.,}{{Bergemann}
  et~al.}{2019}]{bergemann2019}
{Bergemann} M.,  et~al., 2019, \mn@doi [\aap] {10.1051/0004-6361/201935811},
  \href {https://ui.adsabs.harvard.edu/abs/2019A&A...631A..80B} {631, A80}

\bibitem[\protect\citeauthoryear{{Bertin}, {Saglia}  \& {Stiavelli}}{{Bertin}
  et~al.}{1992}]{bertin1992}
{Bertin} G.,  {Saglia} R.~P.,   {Stiavelli} M.,  1992, \mn@doi [\apj]
  {10.1086/170884}, \href
  {https://ui.adsabs.harvard.edu/abs/1992ApJ...384..423B} {384, 423}

\bibitem[\protect\citeauthoryear{{Bradamante}, {Matteucci}  \&
  {D'Ercole}}{{Bradamante} et~al.}{1998}]{bradamante1998}
{Bradamante} F.,  {Matteucci} F.,   {D'Ercole} A.,  1998, \mn@doi [\aap]
  {10.48550/arXiv.astro-ph/9801131}, \href
  {https://ui.adsabs.harvard.edu/abs/1998A&A...337..338B} {337, 338}

\bibitem[\protect\citeauthoryear{{Castrillo} et~al.,}{{Castrillo}
  et~al.}{2021}]{castrillo2021}
{Castrillo} A.,  et~al., 2021, \mn@doi [\mnras] {10.1093/mnras/staa3876}, \href
  {https://ui.adsabs.harvard.edu/abs/2021MNRAS.501.3122C} {501, 3122}

\bibitem[\protect\citeauthoryear{{C{\^o}t{\'e}}, {O'Shea}, {Ritter}, {Herwig}
  \& {Venn}}{{C{\^o}t{\'e}} et~al.}{2017}]{cote2017}
{C{\^o}t{\'e}} B.,  {O'Shea} B.~W.,  {Ritter} C.,  {Herwig} F.,   {Venn} K.~A.,
   2017, \mn@doi [\apj] {10.3847/1538-4357/835/2/128}, \href
  {https://ui.adsabs.harvard.edu/abs/2017ApJ...835..128C} {835, 128}

\bibitem[\protect\citeauthoryear{{Eitner}, {Bergemann}, {Hansen}, {Cescutti},
  {Seitenzahl}, {Larsen}  \& {Plez}}{{Eitner} et~al.}{2020}]{eitner2020}
{Eitner} P.,  {Bergemann} M.,  {Hansen} C.~J.,  {Cescutti} G.,  {Seitenzahl}
  I.~R.,  {Larsen} S.,   {Plez} B.,  2020, \mn@doi [\aap]
  {10.1051/0004-6361/201936603}, \href
  {https://ui.adsabs.harvard.edu/abs/2020A&A...635A..38E} {635, A38}

\bibitem[\protect\citeauthoryear{{Eitner}, {Bergemann}, {Ruiter}, {Avril},
  {Seitenzahl}, {Gent}  \& {C{\^o}t{\'e}}}{{Eitner} et~al.}{2023}]{Eitner2023}
{Eitner} P.,  {Bergemann} M.,  {Ruiter} A.~J.,  {Avril} O.,  {Seitenzahl}
  I.~R.,  {Gent} M.~R.,   {C{\^o}t{\'e}} B.,  2023, \mn@doi [\aap]
  {10.1051/0004-6361/202244286}, \href
  {https://ui.adsabs.harvard.edu/abs/2023A&A...677A.151E} {677, A151}

\bibitem[\protect\citeauthoryear{{Fink}, {Hillebrandt}  \& {R{\"o}pke}}{{Fink}
  et~al.}{2007}]{fink2007}
{Fink} M.,  {Hillebrandt} W.,   {R{\"o}pke} F.~K.,  2007, \mn@doi [\aap]
  {10.1051/0004-6361:20078438}, \href
  {https://ui.adsabs.harvard.edu/abs/2007A&A...476.1133F} {476, 1133}

\bibitem[\protect\citeauthoryear{{Fink}, {R{\"o}pke}, {Hillebrandt},
  {Seitenzahl}, {Sim}  \& {Kromer}}{{Fink} et~al.}{2010}]{fink2010}
{Fink} M.,  {R{\"o}pke} F.~K.,  {Hillebrandt} W.,  {Seitenzahl} I.~R.,  {Sim}
  S.~A.,   {Kromer} M.,  2010, \mn@doi [\aap] {10.1051/0004-6361/200913892},
  \href {https://ui.adsabs.harvard.edu/abs/2010A&A...514A..53F} {514, A53}

\bibitem[\protect\citeauthoryear{{Fl{\"o}rs} et~al.,}{{Fl{\"o}rs}
  et~al.}{2020}]{floers2020}
{Fl{\"o}rs} A.,  et~al., 2020, \mn@doi [\mnras] {10.1093/mnras/stz3013}, \href
  {https://ui.adsabs.harvard.edu/abs/2020MNRAS.491.2902F} {491, 2902}

\bibitem[\protect\citeauthoryear{{Gibson}}{{Gibson}}{1994}]{gibson1994}
{Gibson} B.~K.,  1994, \mn@doi [\jrasc] {10.48550/arXiv.astro-ph/9410031},
  \href {https://ui.adsabs.harvard.edu/abs/1994JRASC..88..383G} {88, 383}

\bibitem[\protect\citeauthoryear{{Gibson}}{{Gibson}}{1997}]{gibson1997}
{Gibson} B.~K.,  1997, \mn@doi [\mnras] {10.1093/mnras/290.3.471}, \href
  {https://ui.adsabs.harvard.edu/abs/1997MNRAS.290..471G} {290, 471}

\bibitem[\protect\citeauthoryear{{Graur} et~al.,}{{Graur}
  et~al.}{2011}]{graur2011}
{Graur} O.,  et~al., 2011, \mn@doi [\mnras] {10.1111/j.1365-2966.2011.19287.x},
  \href {https://ui.adsabs.harvard.edu/abs/2011MNRAS.417..916G} {417, 916}

\bibitem[\protect\citeauthoryear{{Gronow}, {C{\^o}t{\'e}}, {Lach},
  {Seitenzahl}, {Collins}, {Sim}  \& {R{\"o}pke}}{{Gronow}
  et~al.}{2021}]{gronow2021}
{Gronow} S.,  {C{\^o}t{\'e}} B.,  {Lach} F.,  {Seitenzahl} I.~R.,  {Collins}
  C.~E.,  {Sim} S.~A.,   {R{\"o}pke} F.~K.,  2021, \mn@doi [\aap]
  {10.1051/0004-6361/202140881}, \href
  {https://ui.adsabs.harvard.edu/abs/2021A&A...656A..94G} {656, A94}

\bibitem[\protect\citeauthoryear{{Hachisu}, {Kato}, {Saio}  \&
  {Nomoto}}{{Hachisu} et~al.}{2012}]{hachisu2012}
{Hachisu} I.,  {Kato} M.,  {Saio} H.,   {Nomoto} K.,  2012, \mn@doi [\apj]
  {10.1088/0004-637X/744/1/69}, \href
  {https://ui.adsabs.harvard.edu/abs/2012ApJ...744...69H} {744, 69}

\bibitem[\protect\citeauthoryear{{Hartmann}, {Woosley}  \& {El Eid}}{{Hartmann}
  et~al.}{1985}]{hartmann1985}
{Hartmann} D.,  {Woosley} S.~E.,   {El Eid} M.~F.,  1985, \mn@doi [\apj]
  {10.1086/163580}, \href
  {https://ui.adsabs.harvard.edu/abs/1985ApJ...297..837H} {297, 837}

\bibitem[\protect\citeauthoryear{{Hill} et~al.,}{{Hill}
  et~al.}{2019}]{hill2019}
{Hill} V.,  et~al., 2019, \mn@doi [\aap] {10.1051/0004-6361/201833950}, \href
  {https://ui.adsabs.harvard.edu/abs/2019A&A...626A..15H} {626, A15}

\bibitem[\protect\citeauthoryear{{H{\"o}flich}, {Wheeler}  \&
  {Thielemann}}{{H{\"o}flich} et~al.}{1998}]{hoflich1998}
{H{\"o}flich} P.,  {Wheeler} J.~C.,   {Thielemann} F.~K.,  1998, \mn@doi [\apj]
  {10.1086/305327}, \href
  {https://ui.adsabs.harvard.edu/abs/1998ApJ...495..617H} {495, 617}

\bibitem[\protect\citeauthoryear{{Howell} et~al.,}{{Howell}
  et~al.}{2009}]{howell2009}
{Howell} D.~A.,  et~al., 2009, \mn@doi [\apj] {10.1088/0004-637X/691/1/661},
  \href {https://ui.adsabs.harvard.edu/abs/2009ApJ...691..661H} {691, 661}

\bibitem[\protect\citeauthoryear{{Iben} \& {Tutukov}}{{Iben} \&
  {Tutukov}}{1984}]{iben1984}
{Iben} I. J.,  {Tutukov} A.~V.,  1984, \mn@doi [\apjs] {10.1086/190932}, \href
  {https://ui.adsabs.harvard.edu/abs/1984ApJS...54..335I} {54, 335}

\bibitem[\protect\citeauthoryear{{Iwamoto}, {Brachwitz}, {Nomoto}, {Kishimoto},
  {Umeda}, {Hix}  \& {Thielemann}}{{Iwamoto} et~al.}{1999}]{iwamoto1999}
{Iwamoto} K.,  {Brachwitz} F.,  {Nomoto} K.,  {Kishimoto} N.,  {Umeda} H.,
  {Hix} W.~R.,   {Thielemann} F.-K.,  1999, \mn@doi [\apjs] {10.1086/313278},
  \href {https://ui.adsabs.harvard.edu/abs/1999ApJS..125..439I} {125, 439}

\bibitem[\protect\citeauthoryear{{Ji}, {Frebel}, {Simon}  \& {Chiti}}{{Ji}
  et~al.}{2016}]{ji2016}
{Ji} A.~P.,  {Frebel} A.,  {Simon} J.~D.,   {Chiti} A.,  2016, \mn@doi [\apj]
  {10.3847/0004-637X/830/2/93}, \href
  {https://ui.adsabs.harvard.edu/abs/2016ApJ...830...93J} {830, 93}

\bibitem[\protect\citeauthoryear{{Ji} et~al.,}{{Ji} et~al.}{2020}]{ji2020}
{Ji} A.~P.,  et~al., 2020, \mn@doi [\apj] {10.3847/1538-4357/ab6213}, \href
  {https://ui.adsabs.harvard.edu/abs/2020ApJ...889...27J} {889, 27}

\bibitem[\protect\citeauthoryear{{Johnson}, {Kochanek}  \& {Stanek}}{{Johnson}
  et~al.}{2023}]{johnson2023}
{Johnson} J.~W.,  {Kochanek} C.~S.,   {Stanek} K.~Z.,  2023, \mn@doi [\mnras]
  {10.1093/mnras/stad3019}, \href
  {https://ui.adsabs.harvard.edu/abs/2023MNRAS.526.5911J} {526, 5911}

\bibitem[\protect\citeauthoryear{{Jones} et~al.,}{{Jones}
  et~al.}{2019}]{jones2019}
{Jones} S.~W.,  et~al., 2019, \mn@doi [\mnras] {10.1093/mnras/stz536}, \href
  {https://ui.adsabs.harvard.edu/abs/2019MNRAS.485.4287J} {485, 4287}

\bibitem[\protect\citeauthoryear{{Karakas}}{{Karakas}}{2010}]{karakas2010}
{Karakas} A.~I.,  2010, \mn@doi [\mnras] {10.1111/j.1365-2966.2009.16198.x},
  \href {https://ui.adsabs.harvard.edu/abs/2010MNRAS.403.1413K} {403, 1413}

\bibitem[\protect\citeauthoryear{{Keegans} et~al.,}{{Keegans}
  et~al.}{2023}]{keegans2023}
{Keegans} J.~D.,  et~al., 2023, \mn@doi [\apjs] {10.3847/1538-4365/ace102},
  \href {https://ui.adsabs.harvard.edu/abs/2023ApJS..268....8K} {268, 8}

\bibitem[\protect\citeauthoryear{{Kennicutt}}{{Kennicutt}}{1998}]{kennicutt1998}
{Kennicutt} Robert~C. J.,  1998, \mn@doi [\apj] {10.1086/305588}, \href
  {https://ui.adsabs.harvard.edu/abs/1998ApJ...498..541K} {498, 541}

\bibitem[\protect\citeauthoryear{{Kirby}, {Xie}, {Guo}, {Kovalev}  \&
  {Bergemann}}{{Kirby} et~al.}{2018}]{kirby2018}
{Kirby} E.~N.,  {Xie} J.~L.,  {Guo} R.,  {Kovalev} M.,   {Bergemann} M.,  2018,
  \mn@doi [\apjs] {10.3847/1538-4365/aac952}, \href
  {https://ui.adsabs.harvard.edu/abs/2018ApJS..237...18K} {237, 18}

\bibitem[\protect\citeauthoryear{{Kirby} et~al.,}{{Kirby}
  et~al.}{2019}]{kirby2019}
{Kirby} E.~N.,  et~al., 2019, \mn@doi [\apj] {10.3847/1538-4357/ab2c02}, \href
  {https://ui.adsabs.harvard.edu/abs/2019ApJ...881...45K} {881, 45}

\bibitem[\protect\citeauthoryear{{Kobayashi}, {Umeda}, {Nomoto}, {Tominaga}  \&
  {Ohkubo}}{{Kobayashi} et~al.}{2006}]{kobayashi2006}
{Kobayashi} C.,  {Umeda} H.,  {Nomoto} K.,  {Tominaga} N.,   {Ohkubo} T.,
  2006, \mn@doi [\apj] {10.1086/508914}, \href
  {https://ui.adsabs.harvard.edu/abs/2006ApJ...653.1145K} {653, 1145}

\bibitem[\protect\citeauthoryear{{Kobayashi}, {Karakas}  \&
  {Lugaro}}{{Kobayashi} et~al.}{2020a}]{kobayashi2020}
{Kobayashi} C.,  {Karakas} A.~I.,   {Lugaro} M.,  2020a, \mn@doi [\apj]
  {10.3847/1538-4357/abae65}, \href
  {https://ui.adsabs.harvard.edu/abs/2020ApJ...900..179K} {900, 179}

\bibitem[\protect\citeauthoryear{{Kobayashi}, {Karakas}  \&
  {Lugaro}}{{Kobayashi} et~al.}{2020b}]{kobayashi2020b}
{Kobayashi} C.,  {Karakas} A.~I.,   {Lugaro} M.,  2020b, \mn@doi [\apj]
  {10.3847/1538-4357/abae65}, \href
  {https://ui.adsabs.harvard.edu/abs/2020ApJ...900..179K} {900, 179}

\bibitem[\protect\citeauthoryear{{Kroupa}}{{Kroupa}}{2001}]{kroupa2001}
{Kroupa} P.,  2001, \mn@doi [\mnras] {10.1046/j.1365-8711.2001.04022.x}, \href
  {https://ui.adsabs.harvard.edu/abs/2001MNRAS.322..231K} {322, 231}

\bibitem[\protect\citeauthoryear{{Lach}, {R{\"o}pke}, {Seitenzahl}, {Cot{\'e}},
  {Gronow}  \& {Ruiter}}{{Lach} et~al.}{2020}]{lach2020}
{Lach} F.,  {R{\"o}pke} F.~K.,  {Seitenzahl} I.~R.,  {Cot{\'e}} B.,  {Gronow}
  S.,   {Ruiter} A.~J.,  2020, \mn@doi [\aap] {10.1051/0004-6361/202038721},
  \href {https://ui.adsabs.harvard.edu/abs/2020A&A...644A.118L} {644, A118}

\bibitem[\protect\citeauthoryear{{Lanfranchi} \& {Matteucci}}{{Lanfranchi} \&
  {Matteucci}}{2004}]{lanfranchi2004}
{Lanfranchi} G.~A.,  {Matteucci} F.,  2004, \mn@doi [\mnras]
  {10.1111/j.1365-2966.2004.07877.x}, \href
  {https://ui.adsabs.harvard.edu/abs/2004MNRAS.351.1338L} {351, 1338}

\bibitem[\protect\citeauthoryear{{Leung}, {Chu}  \& {Lin}}{{Leung}
  et~al.}{2015}]{leung2015}
{Leung} S.~C.,  {Chu} M.~C.,   {Lin} L.~M.,  2015, \mn@doi [\mnras]
  {10.1093/mnras/stv1923}, \href
  {https://ui.adsabs.harvard.edu/abs/2015MNRAS.454.1238L} {454, 1238}

\bibitem[\protect\citeauthoryear{{Li} et~al.,}{{Li} et~al.}{2022}]{li2022}
{Li} H.,  et~al., 2022, \mn@doi [\apj] {10.3847/1538-4357/ac6514}, \href
  {https://ui.adsabs.harvard.edu/abs/2022ApJ...931..147L} {931, 147}

\bibitem[\protect\citeauthoryear{{Limongi} \& {Chieffi}}{{Limongi} \&
  {Chieffi}}{2018}]{limongi2018}
{Limongi} M.,  {Chieffi} A.,  2018, \mn@doi [\apjs] {10.3847/1538-4365/aacb24},
  \href {https://ui.adsabs.harvard.edu/abs/2018ApJS..237...13L} {237, 13}

\bibitem[\protect\citeauthoryear{{Maoz} \& {Badenes}}{{Maoz} \&
  {Badenes}}{2010}]{maoz2010}
{Maoz} D.,  {Badenes} C.,  2010, \mn@doi [\mnras]
  {10.1111/j.1365-2966.2010.16988.x}, \href
  {https://ui.adsabs.harvard.edu/abs/2010MNRAS.407.1314M} {407, 1314}

\bibitem[\protect\citeauthoryear{{Maoz} \& {Graur}}{{Maoz} \&
  {Graur}}{2017}]{maoz2017}
{Maoz} D.,  {Graur} O.,  2017, \mn@doi [\apj] {10.3847/1538-4357/aa8b6e}, \href
  {https://ui.adsabs.harvard.edu/abs/2017ApJ...848...25M} {848, 25}

\bibitem[\protect\citeauthoryear{{Maoz}, {Mannucci}  \& {Nelemans}}{{Maoz}
  et~al.}{2014}]{maoz2014}
{Maoz} D.,  {Mannucci} F.,   {Nelemans} G.,  2014, \mn@doi [\araa]
  {10.1146/annurev-astro-082812-141031}, \href
  {https://ui.adsabs.harvard.edu/abs/2014ARA&A..52..107M} {52, 107}

\bibitem[\protect\citeauthoryear{{Matteucci}}{{Matteucci}}{2012}]{matteucci2012}
{Matteucci} F.,  2012, {Chemical Evolution of Galaxies},
  \mn@doi{10.1007/978-3-642-22491-1.
}

\bibitem[\protect\citeauthoryear{{Matteucci} \& {Greggio}}{{Matteucci} \&
  {Greggio}}{1986}]{matteucci1986}
{Matteucci} F.,  {Greggio} L.,  1986, \aap, \href
  {https://ui.adsabs.harvard.edu/abs/1986A&A...154..279M} {154, 279}

\bibitem[\protect\citeauthoryear{{Matteucci} \& {Tornambe}}{{Matteucci} \&
  {Tornambe}}{1985}]{matteucci1985}
{Matteucci} F.,  {Tornambe} A.,  1985, \aap, \href
  {https://ui.adsabs.harvard.edu/abs/1985A&A...142...13M} {142, 13}

\bibitem[\protect\citeauthoryear{{McWilliam}, {Piro}, {Badenes}  \&
  {Bravo}}{{McWilliam} et~al.}{2018}]{mcwilliam2018}
{McWilliam} A.,  {Piro} A.~L.,  {Badenes} C.,   {Bravo} E.,  2018, \mn@doi
  [\apj] {10.3847/1538-4357/aab772}, \href
  {https://ui.adsabs.harvard.edu/abs/2018ApJ...857...97M} {857, 97}

\bibitem[\protect\citeauthoryear{{Miles}, {Townsley}, {Shen}, {Timmes}  \&
  {Moore}}{{Miles} et~al.}{2019}]{miles2019}
{Miles} B.~J.,  {Townsley} D.~M.,  {Shen} K.~J.,  {Timmes} F.~X.,   {Moore} K.,
   2019, \mn@doi [\apj] {10.3847/1538-4357/aaf8a5}, \href
  {https://ui.adsabs.harvard.edu/abs/2019ApJ...871..154M} {871, 154}

\bibitem[\protect\citeauthoryear{{Nissen}, {Amarsi}, {Sk{\'u}lad{\'o}ttir}  \&
  {Schuster}}{{Nissen} et~al.}{2024}]{nissen2024}
{Nissen} P.~E.,  {Amarsi} A.~M.,  {Sk{\'u}lad{\'o}ttir} {\'A}.,   {Schuster}
  W.~J.,  2024, \mn@doi [\aap] {10.1051/0004-6361/202348392}, \href
  {https://ui.adsabs.harvard.edu/abs/2024A&A...682A.116N} {682, A116}

\bibitem[\protect\citeauthoryear{{Nomoto}, {Thielemann}  \& {Yokoi}}{{Nomoto}
  et~al.}{1984}]{nomoto1984}
{Nomoto} K.,  {Thielemann} F.~K.,   {Yokoi} K.,  1984, \mn@doi [\apj]
  {10.1086/162639}, \href
  {https://ui.adsabs.harvard.edu/abs/1984ApJ...286..644N} {286, 644}

\bibitem[\protect\citeauthoryear{{North} et~al.,}{{North}
  et~al.}{2012}]{north2012}
{North} P.,  et~al., 2012, \mn@doi [\aap] {10.1051/0004-6361/201118636}, \href
  {https://ui.adsabs.harvard.edu/abs/2012A&A...541A..45N} {541, A45}

\bibitem[\protect\citeauthoryear{{Palla}}{{Palla}}{2021}]{palla2021}
{Palla} M.,  2021, \mn@doi [\mnras] {10.1093/mnras/stab293}, \href
  {https://ui.adsabs.harvard.edu/abs/2021MNRAS.503.3216P} {503, 3216}

\bibitem[\protect\citeauthoryear{{Pignatari} et~al.,}{{Pignatari}
  et~al.}{2016}]{pignatari2016}
{Pignatari} M.,  et~al., 2016, \mn@doi [\apjs] {10.3847/0067-0049/225/2/24},
  \href {https://ui.adsabs.harvard.edu/abs/2016ApJS..225...24P} {225, 24}

\bibitem[\protect\citeauthoryear{{Pipino} \& {Matteucci}}{{Pipino} \&
  {Matteucci}}{2004}]{pipino2004}
{Pipino} A.,  {Matteucci} F.,  2004, \mn@doi [\mnras]
  {10.1111/j.1365-2966.2004.07268.x}, \href
  {https://ui.adsabs.harvard.edu/abs/2004MNRAS.347..968P} {347, 968}

\bibitem[\protect\citeauthoryear{{Prantzos}, {Abia}, {Limongi}, {Chieffi}  \&
  {Cristallo}}{{Prantzos} et~al.}{2018}]{prantzos2018}
{Prantzos} N.,  {Abia} C.,  {Limongi} M.,  {Chieffi} A.,   {Cristallo} S.,
  2018, \mn@doi [\mnras] {10.1093/mnras/sty316}, \href
  {https://ui.adsabs.harvard.edu/abs/2018MNRAS.476.3432P} {476, 3432}

\bibitem[\protect\citeauthoryear{{Rauscher}}{{Rauscher}}{2020}]{rauscher2020}
{Rauscher} T.,  2020, {Essentials of Nucleosynthesis and Theoretical Nuclear
  Astrophysics}, \mn@doi{10.1088/2514-3433/ab8737.
}

\bibitem[\protect\citeauthoryear{{Ritter}, {C{\^o}t{\'e}}, {Herwig}, {Navarro}
  \& {Fryer}}{{Ritter} et~al.}{2018}]{ritter2018}
{Ritter} C.,  {C{\^o}t{\'e}} B.,  {Herwig} F.,  {Navarro} J.~F.,   {Fryer}
  C.~L.,  2018, \mn@doi [\apjs] {10.3847/1538-4365/aad691}, \href
  {https://ui.adsabs.harvard.edu/abs/2018ApJS..237...42R} {237, 42}

\bibitem[\protect\citeauthoryear{{Romano}, {Karakas}, {Tosi}  \&
  {Matteucci}}{{Romano} et~al.}{2010}]{romano2010}
{Romano} D.,  {Karakas} A.~I.,  {Tosi} M.,   {Matteucci} F.,  2010, \mn@doi
  [\aap] {10.1051/0004-6361/201014483}, \href
  {https://ui.adsabs.harvard.edu/abs/2010A&A...522A..32R} {522, A32}

\bibitem[\protect\citeauthoryear{{Rybizki}, {Just}  \& {Rix}}{{Rybizki}
  et~al.}{2017}]{rybizki2018}
{Rybizki} J.,  {Just} A.,   {Rix} H.-W.,  2017, \mn@doi [\aap]
  {10.1051/0004-6361/201730522}, \href
  {https://ui.adsabs.harvard.edu/abs/2017A&A...605A..59R} {605, A59}

\bibitem[\protect\citeauthoryear{{Sanders}, {Belokurov}  \& {Man}}{{Sanders}
  et~al.}{2021}]{sanders2021}
{Sanders} J.~L.,  {Belokurov} V.,   {Man} K. T.~F.,  2021, \mn@doi [\mnras]
  {10.1093/mnras/stab1951}, \href
  {https://ui.adsabs.harvard.edu/abs/2021MNRAS.506.4321S} {506, 4321}

\bibitem[\protect\citeauthoryear{{Savino}, {de Boer}, {Salaris}  \&
  {Tolstoy}}{{Savino} et~al.}{2018}]{savino2018}
{Savino} A.,  {de Boer} T.~J.~L.,  {Salaris} M.,   {Tolstoy} E.,  2018, \mn@doi
  [\mnras] {10.1093/mnras/sty1954}, \href
  {https://ui.adsabs.harvard.edu/abs/2018MNRAS.480.1587S} {480, 1587}

\bibitem[\protect\citeauthoryear{{Schmidt}}{{Schmidt}}{1959}]{schmidt1959}
{Schmidt} M.,  1959, \mn@doi [\apj] {10.1086/146614}, \href
  {https://ui.adsabs.harvard.edu/abs/1959ApJ...129..243S} {129, 243}

\bibitem[\protect\citeauthoryear{{Seitenzahl} et~al.,}{{Seitenzahl}
  et~al.}{2013}]{seitenzahl2013}
{Seitenzahl} I.~R.,  et~al., 2013, \mn@doi [\mnras] {10.1093/mnras/sts402},
  \href {https://ui.adsabs.harvard.edu/abs/2013MNRAS.429.1156S} {429, 1156}

\bibitem[\protect\citeauthoryear{{Shen}, {Kasen}, {Miles}  \&
  {Townsley}}{{Shen} et~al.}{2018}]{shen2018}
{Shen} K.~J.,  {Kasen} D.,  {Miles} B.~J.,   {Townsley} D.~M.,  2018, \mn@doi
  [\apj] {10.3847/1538-4357/aaa8de}, \href
  {https://ui.adsabs.harvard.edu/abs/2018ApJ...854...52S} {854, 52}

\bibitem[\protect\citeauthoryear{{Smartt}}{{Smartt}}{2009}]{smartt2009}
{Smartt} S.~J.,  2009, \mn@doi [\araa] {10.1146/annurev-astro-082708-101737},
  \href {https://ui.adsabs.harvard.edu/abs/2009ARA&A..47...63S} {47, 63}

\bibitem[\protect\citeauthoryear{{Thielemann}, {Nomoto}  \&
  {Yokoi}}{{Thielemann} et~al.}{1986}]{thielemann1986}
{Thielemann} F.~K.,  {Nomoto} K.,   {Yokoi} K.,  1986, \aap, \href
  {https://ui.adsabs.harvard.edu/abs/1986A&A...158...17T} {158, 17}

\bibitem[\protect\citeauthoryear{{Timmes}}{{Timmes}}{1999}]{timmes1999}
{Timmes} F.~X.,  1999, \mn@doi [\apjs] {10.1086/313257}, \href
  {https://ui.adsabs.harvard.edu/abs/1999ApJS..124..241T} {124, 241}

\bibitem[\protect\citeauthoryear{{Timmes}, {Brown}  \& {Truran}}{{Timmes}
  et~al.}{2003}]{timmes2003}
{Timmes} F.~X.,  {Brown} E.~F.,   {Truran} J.~W.,  2003, \mn@doi [\apjl]
  {10.1086/376721}, \href
  {https://ui.adsabs.harvard.edu/abs/2003ApJ...590L..83T} {590, L83}

\bibitem[\protect\citeauthoryear{{Tinsley}}{{Tinsley}}{1979}]{tinsley1979}
{Tinsley} B.~M.,  1979, \mn@doi [\apj] {10.1086/157039}, \href
  {https://ui.adsabs.harvard.edu/abs/1979ApJ...229.1046T} {229, 1046}

\bibitem[\protect\citeauthoryear{{Townsley}, {Jackson}, {Calder}, {Chamulak},
  {Brown}  \& {Timmes}}{{Townsley} et~al.}{2009}]{townsley2009}
{Townsley} D.~M.,  {Jackson} A.~P.,  {Calder} A.~C.,  {Chamulak} D.~A.,
  {Brown} E.~F.,   {Timmes} F.~X.,  2009, \mn@doi [\apj]
  {10.1088/0004-637X/701/2/1582}, \href
  {https://ui.adsabs.harvard.edu/abs/2009ApJ...701.1582T} {701, 1582}

\bibitem[\protect\citeauthoryear{{Townsley}, {Miles}, {Timmes}, {Calder}  \&
  {Brown}}{{Townsley} et~al.}{2016}]{townsley2016}
{Townsley} D.~M.,  {Miles} B.~J.,  {Timmes} F.~X.,  {Calder} A.~C.,   {Brown}
  E.~F.,  2016, \mn@doi [\apjs] {10.3847/0067-0049/225/1/3}, \href
  {https://ui.adsabs.harvard.edu/abs/2016ApJS..225....3T} {225, 3}

\bibitem[\protect\citeauthoryear{{Travaglio}, {Hillebrandt}, {Reinecke}  \&
  {Thielemann}}{{Travaglio} et~al.}{2004}]{travaglio2004}
{Travaglio} C.,  {Hillebrandt} W.,  {Reinecke} M.,   {Thielemann} F.~K.,  2004,
  \mn@doi [\aap] {10.1051/0004-6361:20041108}, \href
  {https://ui.adsabs.harvard.edu/abs/2004A&A...425.1029T} {425, 1029}

\bibitem[\protect\citeauthoryear{{Vincenzo} \& {Kobayashi}}{{Vincenzo} \&
  {Kobayashi}}{2018a}]{vincenzo2018a}
{Vincenzo} F.,  {Kobayashi} C.,  2018a, \mn@doi [\mnras]
  {10.1093/mnras/sty1047}, \href
  {https://ui.adsabs.harvard.edu/abs/2018MNRAS.478..155V} {478, 155}

\bibitem[\protect\citeauthoryear{{Vincenzo} \& {Kobayashi}}{{Vincenzo} \&
  {Kobayashi}}{2018b}]{vincenzo2018b}
{Vincenzo} F.,  {Kobayashi} C.,  2018b, \mn@doi [\aap]
  {10.1051/0004-6361/201732395}, \href
  {https://ui.adsabs.harvard.edu/abs/2018A&A...610L..16V} {610, L16}

\bibitem[\protect\citeauthoryear{{Vincenzo}, {Matteucci}, {de Boer}, {Cignoni}
  \& {Tosi}}{{Vincenzo} et~al.}{2016}]{vincenzo2016}
{Vincenzo} F.,  {Matteucci} F.,  {de Boer} T.~J.~L.,  {Cignoni} M.,   {Tosi}
  M.,  2016, \mn@doi [\mnras] {10.1093/mnras/stw1145}, \href
  {https://ui.adsabs.harvard.edu/abs/2016MNRAS.460.2238V} {460, 2238}

\bibitem[\protect\citeauthoryear{{Vincenzo}, {Matteucci}  \&
  {Spitoni}}{{Vincenzo} et~al.}{2017}]{vincenzo2017}
{Vincenzo} F.,  {Matteucci} F.,   {Spitoni} E.,  2017, \mn@doi [\mnras]
  {10.1093/mnras/stw3369}, \href
  {https://ui.adsabs.harvard.edu/abs/2017MNRAS.466.2939V} {466, 2939}

\bibitem[\protect\citeauthoryear{{Webbink}}{{Webbink}}{1984}]{webbink1984}
{Webbink} R.~F.,  1984, \mn@doi [\apj] {10.1086/161701}, \href
  {https://ui.adsabs.harvard.edu/abs/1984ApJ...277..355W} {277, 355}

\bibitem[\protect\citeauthoryear{{Weinberg} et~al.,}{{Weinberg}
  et~al.}{2022}]{weinberg2022}
{Weinberg} D.~H.,  et~al., 2022, \mn@doi [\apjs] {10.3847/1538-4365/ac6028},
  \href {https://ui.adsabs.harvard.edu/abs/2022ApJS..260...32W} {260, 32}

\bibitem[\protect\citeauthoryear{{Weinberg}, {Griffith}, {Johnson}  \&
  {Thompson}}{{Weinberg} et~al.}{2024}]{weinberg2024}
{Weinberg} D.~H.,  {Griffith} E.~J.,  {Johnson} J.~W.,   {Thompson} T.~A.,
  2024, \mn@doi [\apj] {10.3847/1538-4357/ad6313}, \href
  {https://ui.adsabs.harvard.edu/abs/2024ApJ...973..122W} {973, 122}

\bibitem[\protect\citeauthoryear{{Weisz}, {Dolphin}, {Skillman}, {Holtzman},
  {Gilbert}, {Dalcanton}  \& {Williams}}{{Weisz} et~al.}{2014}]{weisz2014}
{Weisz} D.~R.,  {Dolphin} A.~E.,  {Skillman} E.~D.,  {Holtzman} J.,  {Gilbert}
  K.~M.,  {Dalcanton} J.~J.,   {Williams} B.~F.,  2014, \mn@doi [\apj]
  {10.1088/0004-637X/789/2/147}, \href
  {https://ui.adsabs.harvard.edu/abs/2014ApJ...789..147W} {789, 147}

\bibitem[\protect\citeauthoryear{{Whelan} \& {Iben}}{{Whelan} \&
  {Iben}}{1973}]{whelan1973}
{Whelan} J.,  {Iben} Icko J.,  1973, \mn@doi [\apj] {10.1086/152565}, \href
  {https://ui.adsabs.harvard.edu/abs/1973ApJ...186.1007W} {186, 1007}

\bibitem[\protect\citeauthoryear{{Womack}, {Vincenzo}, {Gibson},
  {C{\^o}t{\'e}}, {Pignatari}, {Brinkman}, {Ventura}  \& {Karakas}}{{Womack}
  et~al.}{2023}]{womack2023}
{Womack} K.~A.,  {Vincenzo} F.,  {Gibson} B.~K.,  {C{\^o}t{\'e}} B.,
  {Pignatari} M.,  {Brinkman} H.~E.,  {Ventura} P.,   {Karakas} A.,  2023,
  \mn@doi [\mnras] {10.1093/mnras/stac3180}, \href
  {https://ui.adsabs.harvard.edu/abs/2023MNRAS.518.1543W} {518, 1543}

\bibitem[\protect\citeauthoryear{{Woosley} \& {Kasen}}{{Woosley} \&
  {Kasen}}{2011}]{woosley2011}
{Woosley} S.~E.,  {Kasen} D.,  2011, \mn@doi [\apj]
  {10.1088/0004-637X/734/1/38}, \href
  {https://ui.adsabs.harvard.edu/abs/2011ApJ...734...38W} {734, 38}

\bibitem[\protect\citeauthoryear{{Woosley} \& {Weaver}}{{Woosley} \&
  {Weaver}}{1994}]{woosley1994}
{Woosley} S.~E.,  {Weaver} T.~A.,  1994, \mn@doi [\apj] {10.1086/173813}, \href
  {https://ui.adsabs.harvard.edu/abs/1994ApJ...423..371W} {423, 371}

\bibitem[\protect\citeauthoryear{{de Boer} et~al.,}{{de Boer}
  et~al.}{2011}]{deboer2011}
{de Boer} T.~J.~L.,  et~al., 2011, \mn@doi [\aap]
  {10.1051/0004-6361/201016398}, \href
  {https://ui.adsabs.harvard.edu/abs/2011A&A...528A.119D} {528, A119}

\bibitem[\protect\citeauthoryear{{de los Reyes}, {Kirby}, {Seitenzahl}  \&
  {Shen}}{{de los Reyes} et~al.}{2020}]{delosreyes2020}
{de los Reyes} M. A.~C.,  {Kirby} E.~N.,  {Seitenzahl} I.~R.,   {Shen} K.~J.,
  2020, \mn@doi [\apj] {10.3847/1538-4357/ab736f}, \href
  {https://ui.adsabs.harvard.edu/abs/2020ApJ...891...85D} {891, 85}

\bibitem[\protect\citeauthoryear{{de los Reyes}, {Kirby}, {Ji}  \&
  {Nu{\~n}ez}}{{de los Reyes} et~al.}{2022}]{delosreyes2022}
{de los Reyes} M. A.~C.,  {Kirby} E.~N.,  {Ji} A.~P.,   {Nu{\~n}ez} E.~H.,
  2022, \mn@doi [\apj] {10.3847/1538-4357/ac332b}, \href
  {https://ui.adsabs.harvard.edu/abs/2022ApJ...925...66D} {925, 66}

\makeatother
\end{thebibliography}

\bsp	
\label{lastpage}
\end{document}